\documentclass[sigconf]{acmart}
\usepackage{caption}
\usepackage{subcaption}
\usepackage{graphicx}
\usepackage{booktabs}
\usepackage{balance}
\usepackage[ruled, vlined, linesnumbered, longend]{algorithm2e}
\SetAlFnt{\small}
\SetAlCapNameFnt{\small}

\AtBeginDocument{%
  \providecommand\BibTeX{{%
    \normalfont B\kern-0.5em{\scshape i\kern-0.25em b}\kern-0.8em\TeX}}}

\copyrightyear{2021}
\acmYear{2021}
\setcopyright{acmcopyright}\acmConference[MM '21]{Proceedings of the 29th ACM International Conference on Multimedia}{October 20--24, 2021}{Virtual Event, China}
\acmBooktitle{Proceedings of the 29th ACM International Conference on Multimedia (MM '21), October 20--24, 2021, Virtual Event, China}
\acmPrice{15.00}
\acmDOI{10.1145/3474085.3481543}
\acmISBN{978-1-4503-8651-7/21/10}

\acmSubmissionID{3370}

\begin{document}
\fancyhead{}

\title{Distantly Supervised Semantic Text Detection and Recognition for Broadcast Sports Videos Understanding}

\author{Avijit Shah}
\orcid{0000-0003-2499-3636}
\affiliation{%
  \institution{Yahoo! Research}
  \city{Sunnyvale}
  \state{CA}
  \country{USA}
}
\email{avijit.shah@verizonmedia.com}

\author{Topojoy Biswas}
\affiliation{%
  \institution{Yahoo! Research}
  \city{Sunnyvale}
  \state{CA}
  \country{USA}
}
\email{topojoy@verizonmedia.com}

\author{Sathish Ramadoss}
\affiliation{%
  \institution{Yahoo! Research}
  \city{Sunnyvale}
  \state{CA}
  \country{USA}
}
\email{rsathish@verizonmedia.com}

\author{Deven Santosh Shah}
\authornote{Work performed while at Yahoo! Research}
\orcid{0000-0001-5653-3023}
 \affiliation{%
   \institution{Microsoft}
   \city{Mountain View}
   \state{CA}
   \country{USA}
 }
\email{devenshah@microsoft.com}



\begin{abstract}
Comprehensive understanding of key players and actions in multiplayer sports broadcast videos is a challenging problem. Unlike in news or finance videos, sports videos have limited text. While both action recognition for multiplayer sports and detection of players has seen robust research, understanding contextual text in video frames still remains one of the most impactful avenues of sports video understanding. In this work we study extremely accurate semantic text detection and recognition in sports clocks, and challenges therein. We observe unique properties of sports clocks, which makes it hard to utilize general-purpose pre-trained detectors and recognizers, so that text can be accurately understood to the degree of being used to align to external knowledge. We propose a novel distant supervision technique to automatically build sports clock datasets. Along with suitable data augmentations, combined with any state-of-the-art text detection and recognition model architectures, we extract extremely accurate semantic text. Finally, we share our computational architecture pipeline to scale this system in industrial setting and proposed a robust dataset for the same to validate our results.
\end{abstract}

\ccsdesc[500]{Computing methodologies~Activity recognition and understanding}
\ccsdesc[500]{Computing methodologies~Visual content-based indexing and retrieval}
\ccsdesc[100]{Computing methodologies~Scene understanding}
\ccsdesc[500]{Computing methodologies~Object detection}
\ccsdesc[500]{Computing methodologies~Object recognition}

\keywords{datasets, neural networks, text detection, text recognition, event detection}


\maketitle
\begin{figure*}
     \centering
     \begin{subfigure}[b]{0.64\textwidth}
         \centering
         \captionsetup{size=scriptsize} 
         \includegraphics[width=\textwidth]{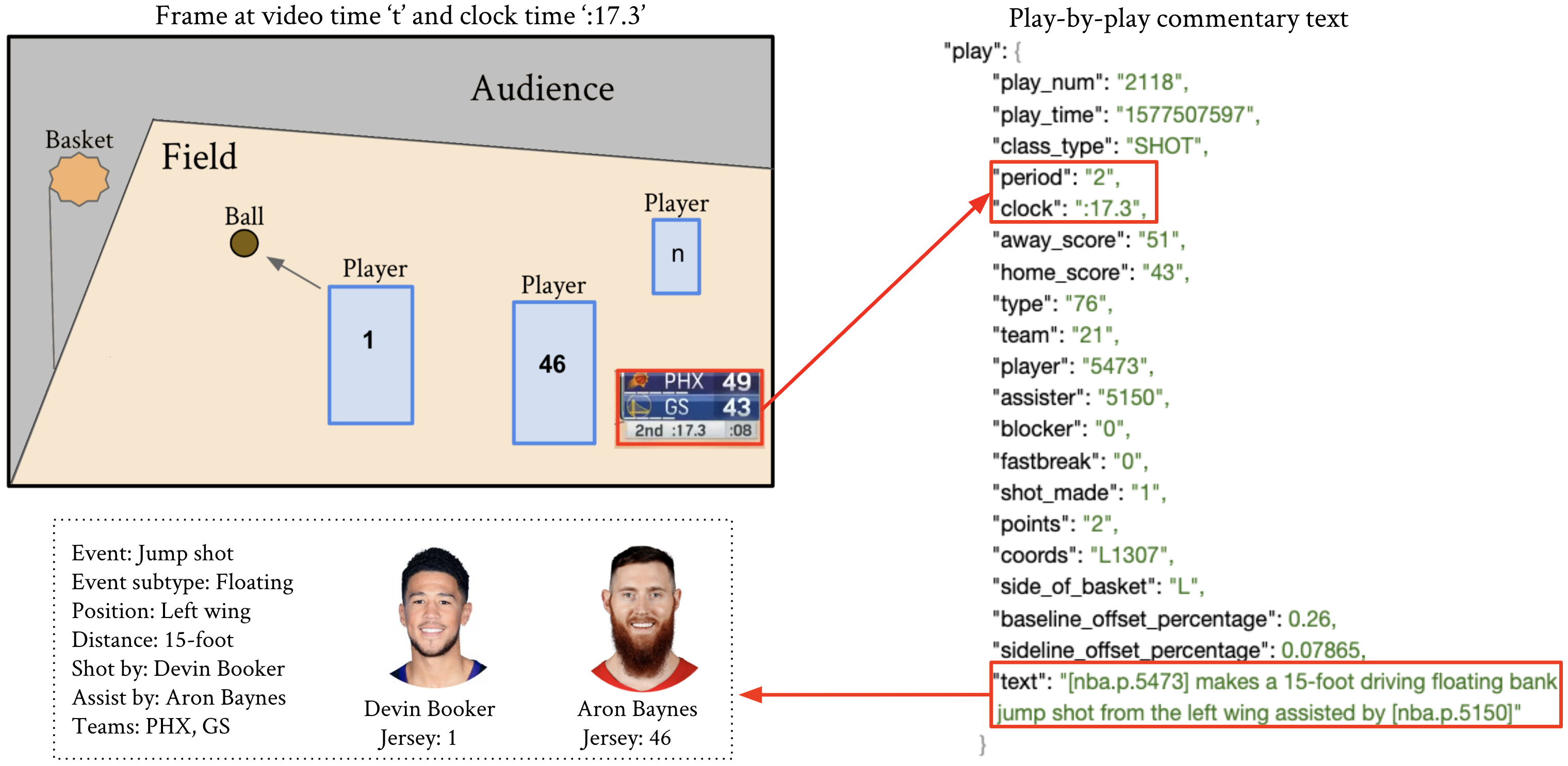}
         \vspace*{-6mm}
         \caption{Comprehensive understanding of video segment by aligning frame with corresponding play-by-play commentary}
         \label{fig:frame-text-alignment}
     \end{subfigure}
     \hfill
     \begin{subfigure}[b]{0.355\textwidth}
        \centering
        \begin{subfigure}[b]{1\textwidth}
            \centering
            \captionsetup{size=scriptsize} 
            \includegraphics[width=\textwidth]{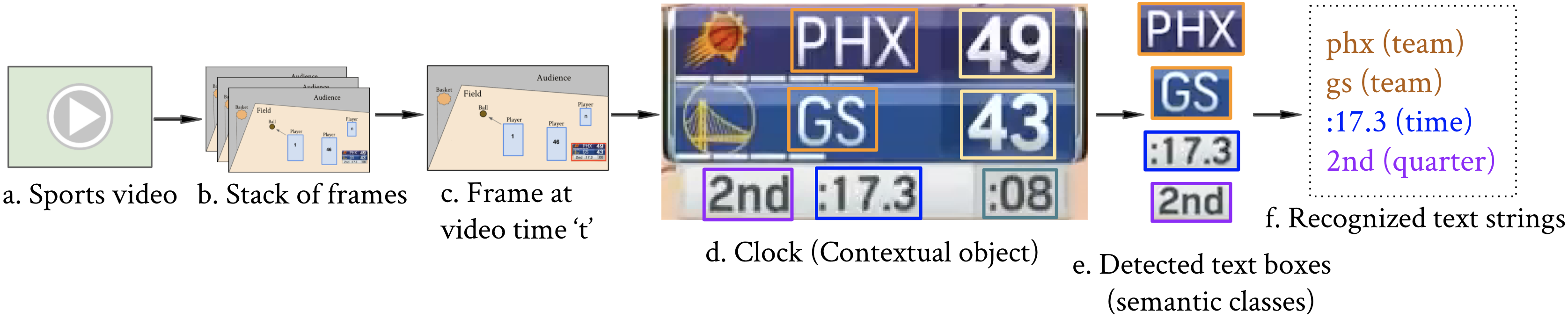}
            \vspace*{-5mm}
            \caption{Process of end-to-end text recognition from sports video using contextual object}
            \label{fig:e-to-e recognition}
        \end{subfigure}
        \begin{subfigure}[b]{0.8\textwidth}
            \centering
            \captionsetup{size=scriptsize} 
            \includegraphics[width=\textwidth]{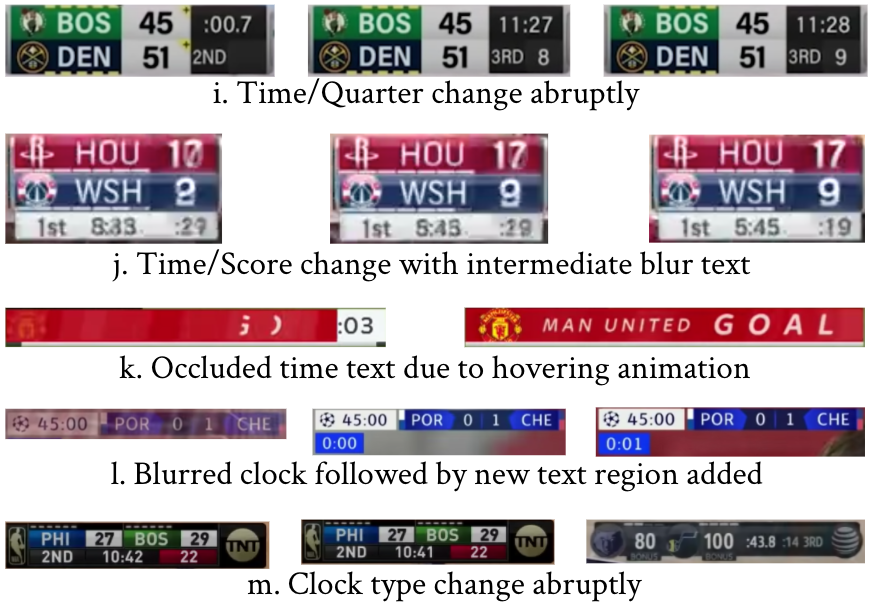}
            \vspace*{-5mm}
            \caption{Scene transitions effects on clocks in sports video highlight}
            \label{fig:Highlight_chop}
        \end{subfigure}
     \end{subfigure}
    \caption{(a) Comprehensive understanding of video segment by aligning frame with corresponding play-by-play commentary using play time and quarter as a composite key. (b) Process of end-to-end text recognition from video frames. (c) Effect of scene transitions in contextual object (clock) in contiguous time interval frames.}
    \label{fig:Main-1}
\vspace*{-4mm}
\end{figure*}
\section{Introduction}
Browsing concise and personalized on-demand sports video highlights, after the game, is a key usage pattern among sports fans. Searching by favourite players, teams and the intent to watch key moments in these videos has been recognized widely\footnote{https://www.thinkwithgoogle.com/consumer-insights/consumer-trends/sports-fans-video-insights-5/}. Enabling such searches or content-based recommendations requires deep indexing of videos at specific moments in time so that the user can be brought to watch the player or moment of intent, without necessarily starting from the beginning of a video. Typically that requires understanding actions happening in video segments and key players in them, apart from participating teams and the date of the game.
\newline
In broadcast videos, key players can be identified either from zoomed-in views just post action (more commonly found in soccer videos, but also in indoor court games, such as basketball), or from the moment a shot is made (by methods like tracking the last human in contact with the ball). Detecting players through both face detection and jersey numbers have been widely studied in recent computer vision conferences. Understanding the context of a sports broadcast video can be achieved through the task of event or action recognition. Doing so accurately at a fine-grained level for multiplayer sports (e.g. professional leagues like \textit{NBA/Basketball, NFL/American football, NHL/ice hockey, Premier League/soccer}) is a challenging task but often necessary in order to support deep video indexing use cases. In several prior research works on action recognition, such as in \cite{playerTrackMurphy,SoccerNet}, the authors discuss methods of generating automatic training data for action recognition by aligning time and quarter information (signifying point in time in a game) available in game clocks to match reports available on the web or in play-by-play records of the game. \textit{Our motivation is to extend such video frame to play-by-play alignment robustly to any freely available sports broadcast highlight video, such that it can directly be used as a method of content understanding at scale, rather than just being a method of gathering training data}.
\newline
Figure \ref{fig:frame-text-alignment} and \ref{fig:e-to-e recognition} show an example of a sample NBA game clock on a broadcast video highlight, and the different text segments in the clock pertaining to team abbreviations ("PHX", "GS"), quarter ("2nd"), and time (":17.3"). Slight misinterpretation of any of these text segments could lead to a potentially wrong play-by-play alignment, or miss a possible alignment altogether. The clock times and quarter transition occurs smoothly, while team names remain same during full game sports broadcast videos, which are usually used for training in the referred works. Hence, correct recognition towards the beginning and end of the game help align the rest of the play-by-plays to the entire game. However, in more abundantly available \textit{short form sports highlights} (typically ranging from 5-15 minutes of video), broadcasters compile several notable plays across the span of the entire game or a season. So, \textit{expected semantic text} like time, quarters and teams in clocks can vary quite frequently. Some videos are compilations from different networks over an entire week or season, so the clocks within the same video or overlaid positions can be different. This makes accurate text detection (such that they can be reliably used for alignment to play-by-plays) a hard problem. Figure \ref{fig:Highlight_chop} shows clocks picked from adjacent video times (frames from 30 or 25 fps videos, which are approximately few frames apart) which experience sudden changes in formatting such as occlusion/blurring, sudden variation in quarter and time, as well as change in clock type in subsequent frames in short game highlights. Such accurate alignments, if done from commonly available short game highlights, can enable deep linking, search and browse use cases across different sports media products at scale in an industrial setting. Text detectors \cite{zhou2017east} and recognizers \cite{shi2015endtoend} which are trained on general purpose datasets tend to not generalize well on domain-specific text and character sequences, with occlusions, occasional blurring and transitions in the varied clock types from different sports. Instead of manually collecting training data for each sport's specific clocks, their text region bounding boxes, and their corresponding texts, we utilize the property that such clock text \textit{come from a finite fixed set of possible strings} (team name abbreviations), as well follow a certain surface form pattern (eg. time consists of digit strings followed by ":" and other digits) as shown in Figure \ref{fig:knowledge-constraints}a. Using this property we develop distant supervision based in-domain text detection and recognition datasets, using which we transfer learn pre-trained detectors and recognizers to the in-domain data. While there has been much study in detecting jersey numbers in order to recognize players via jersey numbers, accurate detection of text in clocks to align them to play-by-play can lead to more fine grained sports video understanding. Thus, our contributions are the following: (1) \textit{we propose a distant supervision approach for in-domain specialized text detection and recognition training}, (2) \textit{a data augmentation method to reduce biases in image sizes and aspect ratios which affect the precision  of text detection and recognition across different sports clocks, without using any domain specific neural architecture}, (3) \textit{we propose a diverse sports clock dataset and comparison with other text detection and recognition datasets to emphasize the nature of the problem} and (4) \textit{this work showcases the practical challenges in making play-by-play alignment work in an industrial setting on highlight videos}. Such multi-modal alignments form the basis for various upstream search and recommendation tasks like in-video search, efficient video scrubbing by deep linking of players and events at specific moments in videos, and content-based recommendation graphs.
\section{Related Work}
\textbf{Identifying key players in the videos:}  In \cite{DBLP:conf/visapp/NadyH21} the authors discuss an end-to-end pipeline of separate player object detectors, text region detection and recognition models, for detecting jersey numbers for basketball and ice hockey. They also propose a dataset for the same. In \cite{Gerke} the authors proposed a dataset focused on recognizing jersey text for soccer and propose a dataset. \cite{Liu} attempted to enrich a Faster RCNN's region proposal with human body part keypoint cues to localize text regions in player jerseys. The authors also propose their own dataset. In \cite{s11042-011-0878-y} the authors approach jersey number localization, in running sports using domain knowledge. 
\newline
\textbf{Broadcast footage text recognition}: In \cite{ref1} authors describe a method of localizing broadcast text region using traditional vision techniques, while \cite{ref2} authors showcase an architecture specific to the task of understanding clock text.
\newline
\textbf{Event detection for contextual information:} Recent works on detecting actions in basketball (NBA) \cite{ramanathan2016detecting}, baseball (MLB) \cite{piergiovanni2018finegrained}  and soccer \cite{SoccerNet} attempt to solve the problem at a coarse-grained level. In \cite{Yu_2018_CVPR} the authors propose an approach to fine-grained captioning though not covering player identification or fine-grained shooting actions for basketball. 
\newline
\textbf{Text detection and recognition:} Text detection model architectures in recent works have ranged from SSD \cite{LiuAESR15} motivated, anchor box based TextBox++ \cite{textbox}, designed for horizontally aligned text. However, like in SSD, the anchor boxes have to be tuned outside regular training, based on the common image and text box sizes in the training dataset. Seglink \cite{seglink} links neighbouring text segments based on spatial relations, helping identify long lines as in Latin. CTPN \cite{ctpn} used an anchor mechanism to predict the location and score of each fixed-width proposal simultaneously, and then connected the sequential proposals by a recurrent neural network. Finally in EAST \cite{zhou2017east}, a fully convolutional network is applied to detect text regions directly without using the steps of candidate aggregation and word partition, and then NMS is used to detect word or line text. In \cite{shi2015endtoend}, CNN-based image representations combines with a Bi-LSTM layer for text recognition. 

\section{Approach}
\label{sec:Approach}
\begin{figure}[htp]
    \centering
    \includegraphics[width=\columnwidth]{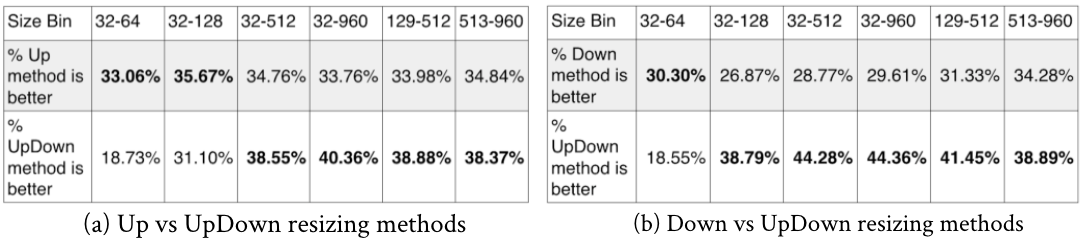}
    \vspace*{-7mm}
    \caption{Aspect ratio distortion comparison}
    \label{fig:ResizeAlgoCompare}
\vspace*{-3mm}
\end{figure}

Accurate semantic text detection and recognition (time left in clock or time elapsed, quarter or half of the game, abbreviation of team names) from sports clocks and matching them to an external knowledge of play-by-play records yields fine-grained player and action detection. For such alignment or linking to external knowledge bases, its critical that the limited pieces of semantic texts are properly understood in the clock. Instead of relying on fine grained image classification (to different teams, or times, as often done in case of jersey number identification of players) or any domain specific neural architecture, or any classical vision/geometric heuristic (for text localization as in \cite{ref1}), we resort to accurate text region detection and text recognition methods (using well used model architectures for maintainability and ease of use in production environments), without getting large sets of humanly labelled sports clock domain training data. Domain-specific knowledge of correctness of detection and recognition can be utilized to mitigate requirement of hand labelled data. Parallely there has been related work on using domain knowledge to regularize model posteriors \cite{hu2016harnessing, hu2018deep}. We formalize domain knowledge of correctness of recognized team names, time, or half/quarter as first-order logical rules and refer to them as Knowledge Constraints (KC). Set of correctly recognized \textit{team names} and \textit{game quarters/halves} form a low cardinality set and \textit{time} comes from a set of known surface forms. We innovate on distilling out correct training data from a pool of noisy instances using KC as a medium for distant supervision. Specifically, we run off the shelf pre-trained text detectors and recognizer on unlabelled cropped clock objects. Predictions from such general-purpose models tend to be inexact, which we correct or reject based on the knowledge constraints to gather high quality in-domain training data without manual labelling. We discuss on KCs in detail in section \ref{sec:knowledge-constraints}.
Our end-to-end process of aligning sports clocks to play-by-play is broken down into the following steps (illustrated in Figure \ref{fig:frame-text-alignment} and \ref{fig:e-to-e recognition}). :
\newline
\textbf{Extracting clock regions from full frames:} For this purpose we train a clock object detector by collecting training samples from different sports highlight videos both internal as well freely available on YouTube. We use semi-automatic methods (detailed in section \ref{section:dataset}) to gather bounding boxes for sports clocks of various sizes, shapes and colors across different sports. We train a single shot object detector \cite{LiuAESR15} with a VGG16 backbone. The SSD bounding box loss consists of localization loss, \begin{math}L_{loc}\end{math} (a smooth L1 loss) which starts with positive samples and regresses the predictions closer to the matched ground-truth boxes and the confidence loss \begin{math}L_{conf}\end{math} (a softmax loss), used for optimizing classification confidence of objects over multiple classes. In our case we resort to only two classes \textit{clock} and \textit{background}, and we increase the importance of the localization loss to 3 times the confidence loss, such as to obtain a tight bounding box around the clock, while not caring so much about the right class prediction.
\newline
\textbf{Semantic text area detection on clock objects:}
Once we extract the clocks from the video frames, we get to the next stage of detecting text segments. Here we adapt the framework of EAST \cite{zhou2017east}, primarily because of its lightweight layers. EAST is a fully convolutional network (FCN) based text detector, which eliminates intermediate steps of candidate proposal, text region formation and word partition. The backbone in EAST is a pre-trained ResNet-50.
\newline
\textbf{Adaptation of text detector to our domain:} We pass the cropped clocks to an in-domain trained (in section \ref{sec:knowledge-constraints} we discuss how we generate in-domain text bounding box ground truths) EAST model, to detect the different text areas (of team, time, quarter). To capture image representations of different resolutions four levels of feature maps, are extracted from the backbone, whose sizes are 1/32, 1/16, 1/8 and 1/4 of the input image, respectively \cite{zhou2017east}. Since the backbone's most granular convolution feature covers 1/32 of the size of an image, and clock cropped images may not come as a multiple of 32, we resort to resizing before passing to the prediction stage. Usually an image is up-scaled or down-scaled to a multiple of 32, \textit{however we amalgamate both techniques for each dimension such that their distortion from their original values are the least, in turn providing less information loss along each axis as well as lesser aspect ratio distortion overall}. In Figure \ref{fig:ResizeAlgoCompare}, we compare relative aspect ratio distortions of just up-scaling or down-scaling with our amalgamation method in different size ranges and find it to achieve fewer distortion cases for our target image sizes of multi-domain sports clocks.

\textbf{Text recognition on detected semantic text areas:} Finally, we crop out text regions inside the clocks and pass them to an in-domain trained text recognizer. We fine tune a CRNN \cite{shi2015endtoend} model with our sports domain text. \cite{shi2015endtoend} combine CNN-based image representations with a Bi-LSTM layer to detect the character sequence with a Connectionist Temporal Classification layer (CTC). 
\newline
\textbf{Adaptation of text recognizer to our domain:} Since we have a good proportion of special characters in our dataset, we train a CRNN model \cite{shi2015endtoend} by combining Synth90k dataset \cite{synth90k1,synth90k2} with our in-domain data. While time and quarter texts have important punctuation (e.g., ":"), they also pose different character sequences for recognition, unavailable in synthesized text. Hence, we include in-domain clock text regions and appropriate strings as shown in Figure \ref{fig:e-to-e recognition} in the training data. All words and characters are converted to lowercase to reduce the complexity of predictions. As for our domain, if the full strings predicted precisely match the knowledge base or the known surface form, we can uniquely identify team names, time or quarter (e.g., "23:21", "1st", "PHX"). 
\newline
 \textbf{Aligning to play-by-play}: Play-by-play commentary can be visualized as text captions or structured data as shown in Figure \ref{fig:frame-text-alignment}, which is a fine-grained description of the game state. If the team names, time, and quarter, as shown in Figure \ref{fig:e-to-e recognition}, in the game clock signifying the game state, can be accurately recognized, then they can be used to uniquely align a group of consecutive video frames (assuming a video is typically more than 1fps) to the play-by-play caption. Such alignments yield a fine-grained understanding of sports broadcast video scenes as shown in Figure \ref{fig:frame-text-alignment}.

\subsection{Distant supervision using Knowledge Constraints (KC)}
\vspace*{-4mm}
\label{sec:knowledge-constraints}
\begin{figure}[htp]
    \centering
    \includegraphics[width=8.5cm]{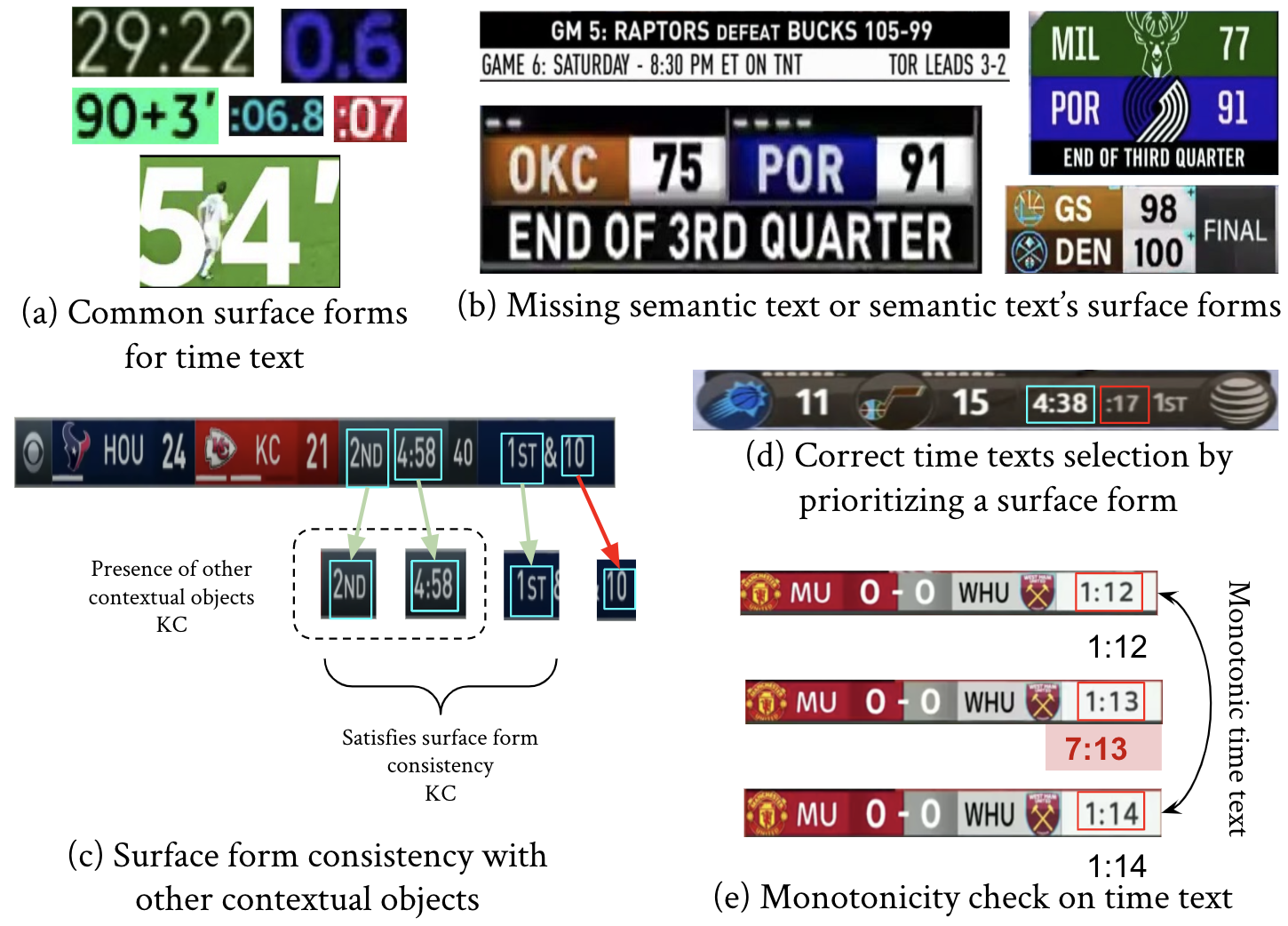}
    \vspace*{-7mm}
    \caption{Distant supervision through KC}
    \label{fig:knowledge-constraints}
\vspace*{-4mm}
\end{figure}
The KC can be broadly classified into the following:
\newline
\textbf{Semantic text surface form check  (KC1)}: Time text on the clocks can be easily recognized by their surface forms: \textit{n\string{1+\string}\textbf{:}n\string{1+\string}} or \textit{n\string{1+\string}\textbf{:}n\string{1+\string}\textbf{.}n\string{1+\string}} as shown in Figure \ref{fig:knowledge-constraints}a. In case of leagues like NBA, NFL, and NHL, time representing quarters have limited string combinations like \textit{"1st", "2nd", "3rd" or "4th"}, while in soccer, times in both play-by-play and the game clocks are \textit{continuous} i.e. they do not reset as the game rolls on to the next half. To ensure that the right contextual object is cropped from the frame, we check for the existence of all semantic text classes in them. For example, in Figure  \ref{fig:knowledge-constraints}b, shown cropped images are very likely to be confused for the contextual object (clock) by any object detection model. Whereas they have some semantic texts in each of them, and none has all semantic texts. Specifically, we select clocks with at least two team texts, one quarter text, and one time text surface form as potential training candidates. Incorrect clocks are used for hard negative mining to improve clock detection. 
\newline
\textbf{Surface form consistencies with the other contextual objects (KC2)}: Some leagues or sports require us to disambiguate further text recognition beyond surface form checks (as there could be multiple candidates), e.g., to disambiguate text representing quarter in NFL, there could be two strings \textit{1st} followed by \textit{10:32} and \textit{2nd} followed by \textit{\& 10} as illustrated in Figure \ref{fig:knowledge-constraints}c, former represents the quarter and time on the clock, and the latter represents players down, with yards remaining. We use the detection of a time surface form and the nearest qualifying quarter text as the right pick. 
\newline
\textbf{Correct time text selection by prioritizing a surface form (KC3)}: It is often observed that there exist multiple candidates for the same semantic text but with different surface forms. For example. as shown in Figure \ref{fig:knowledge-constraints}d, \textit{4:38} and \textit{:17} are potential candidates for time text. We set up the priorities of time text surface forms based on the external knowledge we gathered about the clock.
\newline
\textbf{Temporal consistencies across video frames (KC4)}: Finally, with the presence of surface form consistency and correct time text known, we validate the text recognition mistakes made by a general-purpose model by the continuity of corresponding text in the next frame. Specifically, we convert time text to their corresponding value in seconds or minutes and compare if they are monotonically increasing (in soccer) or decreasing (in NFL, NBA, and NHL) in group of three consecutive frames. The recognized clock sequence time text not following the monotonicity can either be discarded or corrected if predictions in previous and subsequent frames largely comply with monotonicity. An example of a soccer clock sequence is shown in Figure \ref{fig:knowledge-constraints}e.
\vspace*{-4mm}
\subsection{Data augmentation for bias correction}
\label{sec:dataaug}
 We observe that the typical size of images and number of text regions (especially NFL versus other leagues/sports), and aspect ratios (especially NBA versus soccer) vary across sports domains. For generalizing models on all sports domains (e.g., soccer), we collect wide range of different broadcaster/sport clocks across different seasons. We further analyze the size dimensions (height and width of the sports clocks) and styles of clocks by clustering them and find representative samples and cluster population sizes. 
\newline
\textbf{Aspect ratio and size sensitive custom data augmentation} (represented as custom data augmentation in algorithm \ref{algo_main} step 11): Post correction/filtering of the text string and bounding box labels using knowledge constraints mentioned in section \ref{sec:knowledge-constraints}, we calculate the possible range of height and width across different sports clocks (NBA/soccer/NFL/NHL). We found that most variations are present between NBA and soccer clocks (detailed in section \ref{section:dataset}). We break down both clock image dimension ranges, and sliced text box dimension ranges into buckets and measure population density in each of these buckets. We then re-scale each original-sized soccer or NBA clock image to randomly fall in a bucket of the fewer population such that the population densities across buckets normalize. As a result, we observe that a typical NBA clock resized between a min-max range of 0.4x-1.2x of its original size, while for a relatively small soccer clock, the range is 0.7x-1.2x. This fixed range of NBA and soccer clocks shows presence of label bias \cite{Shah_2020}. These data augmentations are carried out in addition to random resizing which is a standard augmentation approach for text detector training.
\newline
\textbf{Labels correction for sport sub-domains:} Taking advantage of the bias present in training set for the text detector and recognizer model, we resize the unlabeled clock images from soccer to the typical sizes of NBA clocks and predict the labels from the NBA sub-domain trained text detector and recognizer. We then filter correct labels by applying knowledge constraints for the target sub-domain (soccer) and then scale back clock images and text box ground truths to original sizes. This is done before the data augmentation described above. Initially  we train for the sub-domain of NBA by using AWS Rekognition as noisy pre-trained model. For other sports domains, we have a choice of AWS Rekognition or our NBA trained model as the pre-trained model. The choice of which pre-trained model to pick (in the wild AWS Rekognition or sub-domain trained NBA model) depends on how minimal the aspect ratio/size sensitive resizing is required such that the out-of-domain detector and recognizer perform well on the new sub-domain (e.g. new sport or league). In section \ref{section:dataset}, we discussed the individual properties of the sub-domain datasets (e.g. NBA, soccer).
\vspace*{-3mm}

\begin{algorithm}[]
\SetAlgoLined
\SetKwInOut{KwIn}{Input}
\SetKwInOut{KwOut}{Output}

\KwIn{Out-of-domain text detector and recognizer $otdr$; knowledge constraints $kc$ = \{KC1, KC2, KC3, KC4\}; sport domains $sd$ =\{NBA, soccer, NHL, NFL\}}
\KwOut{Generalized text detection and recognition models for all $sd$, $itdr$; Clean text detection and recognition training dataset $T$}
$itdr$ $ \leftarrow $ $otdr$\;
Clean text detection and recognition training dataset $T$ $ \leftarrow $ None\;
\For{$sport \leftarrow NBA$ \KwTo NFL}{
$videos$ $ \leftarrow $ set of videos from $sport$\;
\For{$i \leftarrow 1$ \KwTo number of videos}{
$clocks$  $ \leftarrow $ (cropped clocks from $videos[i]$, frame ids)\;
clocks with noisy text boxes and noisy recognized text $clocks_{n}$ $ \leftarrow $ $itdr ( clocks )$\; 
clocks with clean text boxes and clean recognized text $clocks_{c}$  $ \leftarrow $ $kc (clocks_{n})$\;
$T$ $ \leftarrow $ $T\cup clocks_{c}$
}
Do custom data augmentation on $T$\;
Do stratified sampling across different clock types on $T$\;
$itdr$ $ \leftarrow $ subdomain adaptive learning update:\
Transfer learn text detection model on all text boxes from clocks in $T$;
Transfer learn text recognizer model on text regions pertaining to semantic classes (time, quarter, team names) mixed with Synth90k data\;
}
Return $itdr$ and $T$\;
 \caption{Text detection and recognition models learning algorithm for all sports domains}\label{algo_main}
\end{algorithm}    
\vspace{-5mm}

\section{Dataset}
\label{section:dataset}
We collect an extensive collection of sports clocks across NBA, soccer, NFL, and NHL sports and broadcasting networks varying in look and feel, sizes, and fonts. For each such clock, we collect the text bounding boxes encompassing the different types of semantic text {like time left in the quarter, team names abbreviations and score}, and the recognized strings. However, our focus is on team names, time, and quarter texts.
\newline
\textbf{Crawling and clock bounding box detection:} We crawled around 1,500 videos, which are short highlights (30 seconds - 2 min) of 720p resolution from nba.com\footnote{https://www.nba.com/sitemap\_video\_*.xml}\footnote{https://www.nba.com/sitemap\_index.xml} or longer highlights videos averaging to 10 minutes in duration from official NBA and soccer league channels from Youtube\footnote{https://www.youtube.com/user/NBA}\footnote{https://www.youtube.com/channel/UCqZQlzSHbVJrwrn5XvzrzcA}\footnote{https://www.youtube.com/c/BRFootball/featured}\footnote{https://www.youtube.com/c/beinsportsusa/featured}. We use CVAT \footnote{https://github.com/openvinotoolkit/cvat} tool to semi-automatically annotate clock object bounding boxes of around 40,000 frames of the above videos. CVAT's object tracking feature reduces manual effort in annotating clocks which are relatively static in their spatial position across the span of the videos. Finally, using the in-domain trained SSD \cite{LiuAESR15} model as stated in section \ref{sec:Approach}, we predict clock bounding boxes and crop nearly 80,049 clocks from the pool of downloaded videos.
\newline
\textbf{Text detection and recognition ground truth generation using knowledge constraints:}
As mentioned in section \ref{sec:knowledge-constraints}, we resort to knowledge constraints to filter quality ground truth text bounding boxes and recognition semantic text strings from noisy predictions of pre-trained models. We use AWS Rekognition \footnote{https://aws.amazon.com/rekognition/} as our pre-trained models both for detection and recognition. Figure \ref{fig:AWSModel_errors} shows detection and recognition errors made by the pre-trained model and results of the in-domain transfer learned model on training data corrected by knowledge constraints.
\newline
\begin{figure}[htp]
    \centering
    \includegraphics[width=8.5cm]{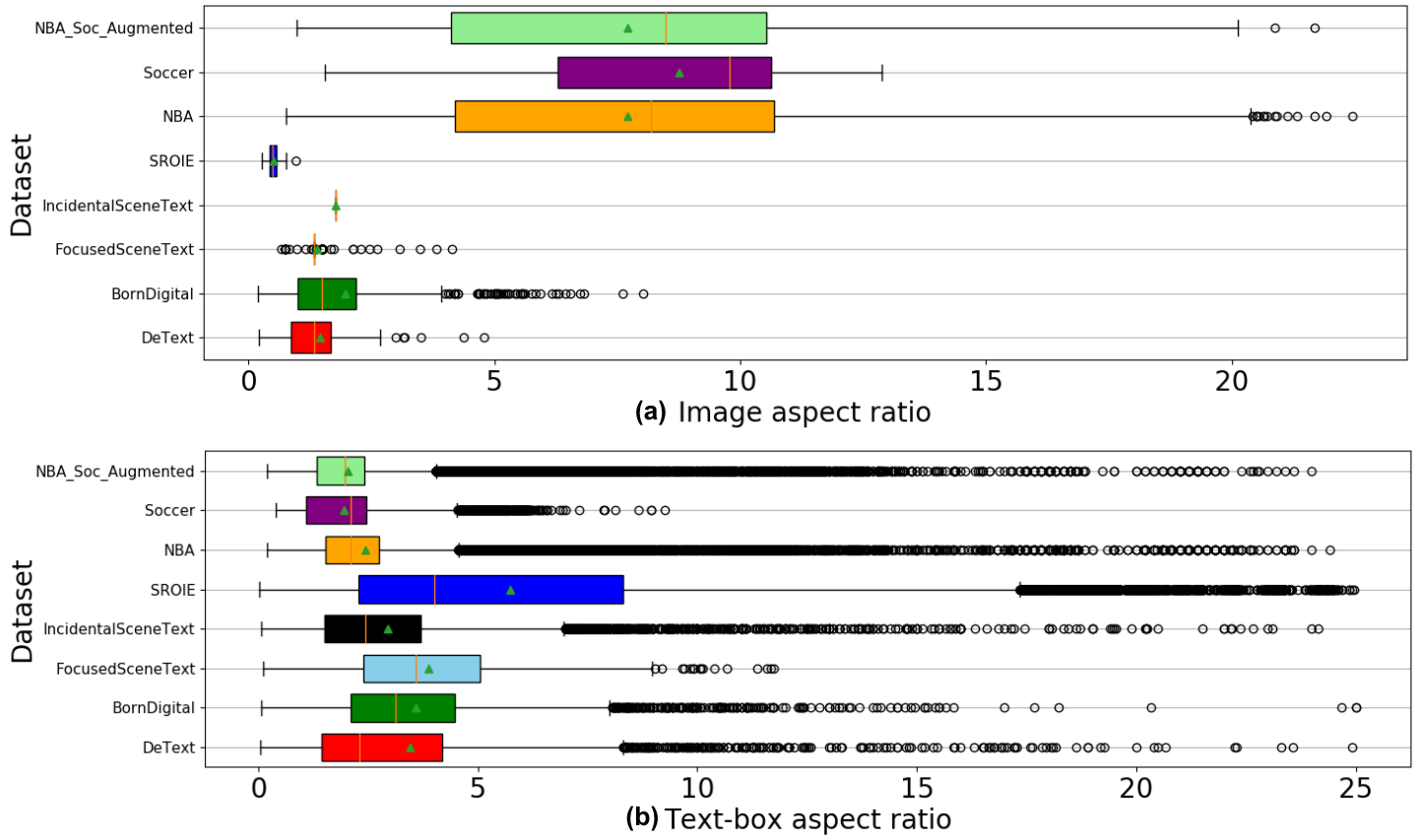}
    \vspace*{-6mm}
    \caption{Text box and image aspect ratio comparison}
    \vspace*{-2mm}
    \label{fig:AspecRationComparison}
\end{figure}
We generate cropped clocks for other sports like soccer, American football (NFL), and ice hockey (NHL) in the same manner. We use the in-domain trained model on the NBA clocks as the pre-trained model for other sports since it generalizes better to other sports clocks than in-the-wild trained text detectors. But we still filter predictions using knowledge constraints for the target sports.
\begin{figure}[htp]
    \centering
    \includegraphics[width=8.5cm]{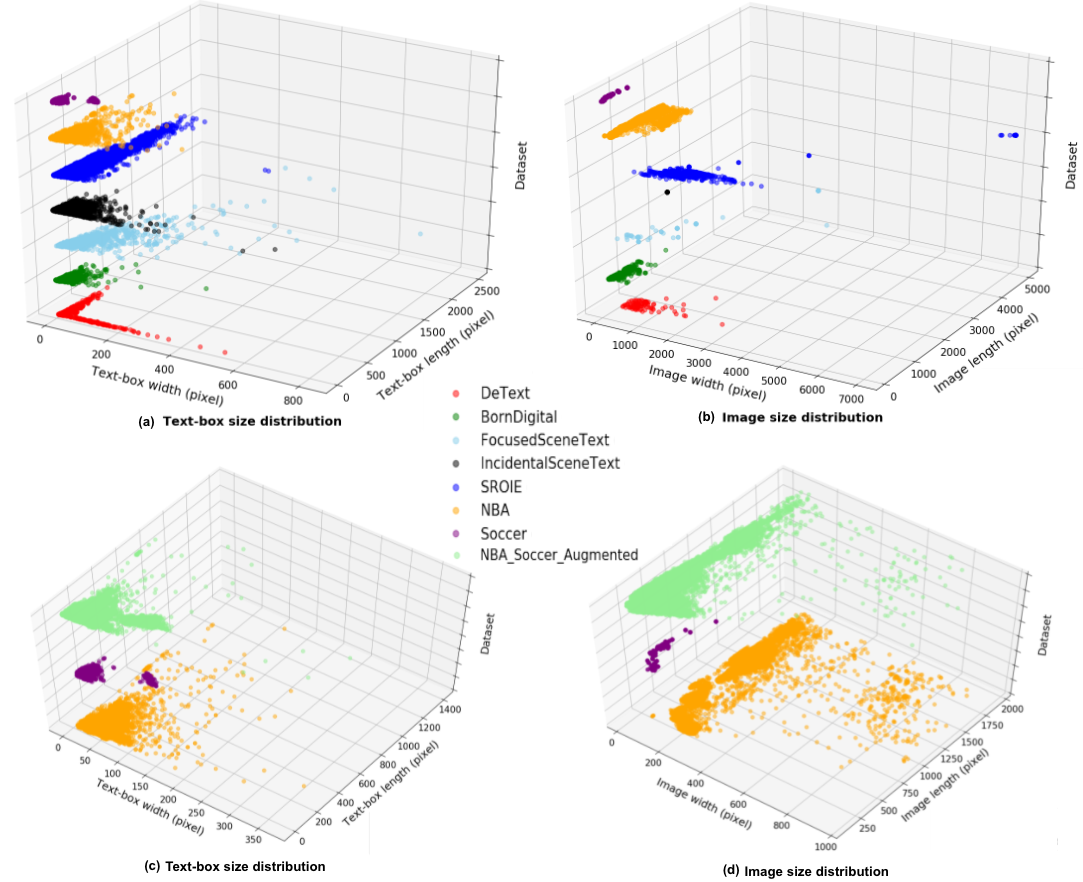}
    \vspace*{-6mm}
    \caption{Distribution of image and text size across datasets}
    \vspace*{-2mm}
    \label{fig:ALLimageSize}
\end{figure}

\textbf{Research dataset and origins:}
We compare our dataset\footnote{will be made available at https://webscope.sandbox.yahoo.com/} with the dataset properties of ICDAR challenges appearing from different domains. Born-digital, ICDAR 2015 \cite{10.1109/ICDAR.2013.221, inproceedings}, the dataset contains low-resolution images with texts digitally created on them, taken from web content. Focused Scene Text dataset in ICDAR 2015 \cite{10.1007/s10032-004-0134-3} has captured images from majorly horizontally aligned scenes text without significant blurring. Incidental Scene Text dataset in ICDAR 2015 \cite{7333942} contains casually in-the-wild captured images, including blur and text alignments. DeText dataset in ICDAR 2017 \cite{8270166} has images specifically about graphs, tables, and charts with text focused on biomedical literature. SROIE, ICDAR 2019 \cite{8395172, 10.1109/ICDAR.2013.221, 7333942} dataset is generated by combining data from different resources focused on reading text from images obtained by scanning receipts and invoices. 

\begin{table}
\caption{Datasets text properties and our dataset size}
\vspace{-4mm}
\begin{subtable}{\columnwidth}
  \captionsetup{size=scriptsize}
  \caption{Properties comparison between different datasets. \textsuperscript{*} represents strings which are composed of only digits, punctuations and space characters}
  \vspace{-2mm}
  \resizebox{\columnwidth}{!}{%
  \begin{tabular}{cccccccc}
    \toprule
    Dataset& Textboxes& Words\textsuperscript{*} & Characters &  Characters\textsuperscript{*} &  words\textsuperscript{*} (\%)& characters\textsuperscript{*} (\%)\\
    \midrule
     IncidentalSceneText& 11875 & 222 & 46070  & 1418 & 1.86(\%) &  3.07(\%)\\
     FocusedSceneText& 849 &  40 & 4784 & 296 & 4.71 (\%) & 6.81 (\%)\\
     BornDigital& 4204  & 403  & 21317  & 2164  & 9.58 (\%) & 10.15 (\%)\\
     SROIE& 45279 & 14561  & 525183 & 222216 & 32.16 (\%) & 42.31 (\%)\\
     DeText& 2314  & 1026  & 17199 & 5930 & 44.33 (\%) & 34.47 (\%)\\
     NBA&  397807   & 165527  & 1414348 & 517046 & 41.61 (\%) & 36.56 (\%)\\
     Soccer& 66792  & 32471  & 188006 & 87112 & 48.62(\%) & 46.33 (\%)\\
  \bottomrule
\end{tabular}%
}
\label{tab:PropertyComp}
\end{subtable}

\begin{subtable}{\columnwidth}
\captionsetup{size=scriptsize}
\caption{Total clocks semantic objects samples in our datasets. Knowledge constraints based cleaning, KC1 and KC4 as mentioned in section \ref{sec:knowledge-constraints}. KC4 comprises of KC2 and KC3}
\vspace{-2mm}
\resizebox{\columnwidth}{!}{%
\begin{tabular}{ccccccccc}
\toprule
Dataset                 & Clock Styles & Aug. NBA (subset) & Aug. soccer & Noisy & KC1 & KC4 & Clean \\
\midrule
NBA                        & 20           & -                              & -                        & 80049 & 6677 & 7389 & 65983                          \\
Soccer                           & 10           & -                              & -                        & 20437 & 378  & 1221 & 18838                          \\
Generalised & 30            & 29117                          & 52372                    & -     & - & -         & 81489 \\
\bottomrule
\end{tabular}%
}
\label{tab:Dataset Size}
\end{subtable}
\label{tab:Dataset and Property}
\vspace{-4mm}
\end{table}

In Figure \ref{fig:AspecRationComparison} we compare image aspect ratio and text box ground truth's aspect ratio across different datasets. We observe that our sports clock sub-domains (NBA, soccer) have considerably higher image aspect ratios (q1-q3 quartiles in the box plots) as compared to other ICDAR datasets, while in comparison our text box aspect ratios are relatively small among the same datasets. We see considerable relative variation in image \text{aspect ratios} among NBA and soccer, while their text box aspect ratios are almost comparable. In figures \ref{fig:ALLimageSize}a and \ref{fig:ALLimageSize}b we find considerable variation in \textit{image sizes} and \textit{text box sizes} between our in-domain datasets (NBA and soccer) versus various ICDAR datasets. In addition, Figure \ref{fig:ALLimageSize}c and Figure \ref{fig:ALLimageSize}d show how variation in NBA and soccer text box and image sizes are handled in augmented dataset. This justifies our need for large in-domain datasets as well as handling variations in image sizes in sub-domains and that's why specifically aspect ratio preserving image resizing strategies discussed under data augmentations in section \ref{sec:dataaug} performs well. As seen in Table \ref{tab:PropertyComp}, our sports clock datasets from NBA and soccer contain texts which are richer in numbers and special characters (many being only numbers and special characters). Text in our dataset are not majorly blurred or are arbitrarily aligned (as in real scene images). Finally in Table \ref{tab:Dataset Size} we report the dataset sizes for sports clocks. In the columns KC1 and KC4 we report number of noisy examples which are rejected, as we sequentially apply each knowledge constraint.


\begin{figure*}
     \centering
     \begin{subfigure}[b]{0.65\textwidth}
         \centering
         \captionsetup{size=scriptsize}
         \includegraphics[width=\textwidth]{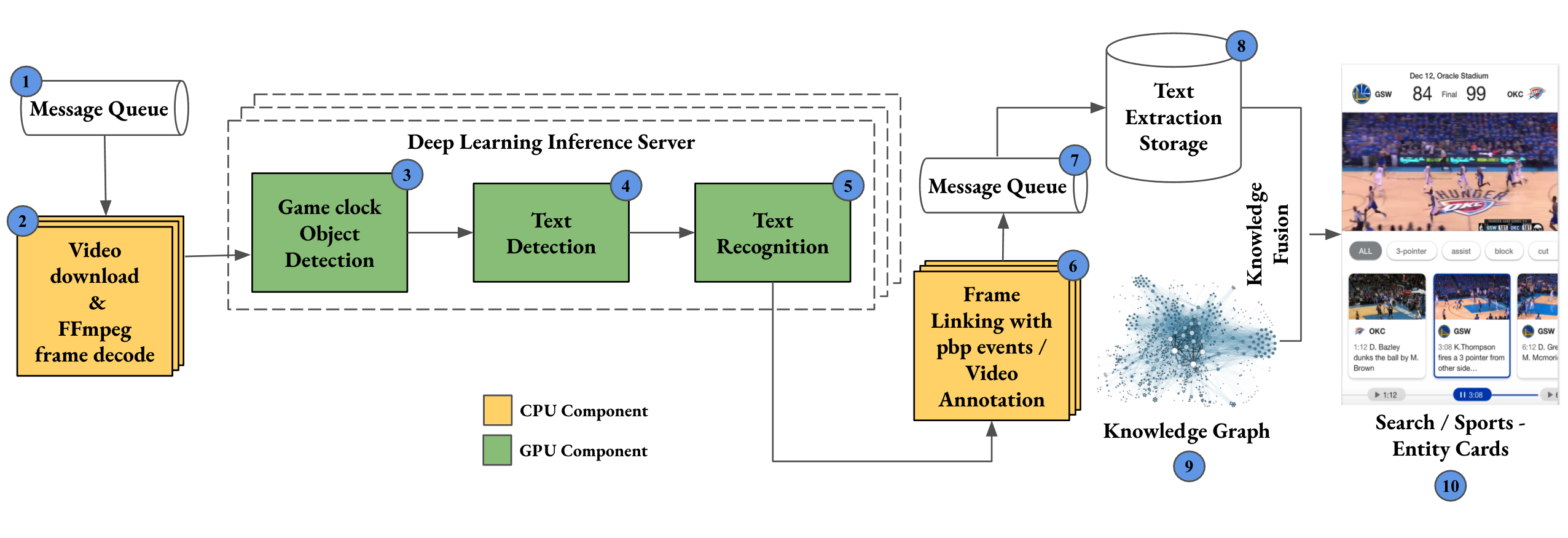}
         \vspace*{-8mm}
         \caption{Computational Architecture}
         \label{fig:computational-architecture}
     \end{subfigure}
     \hfill
     \begin{subfigure}[b]{0.29\textwidth}
        \centering
        \captionsetup{size=scriptsize}
        \includegraphics[width=\textwidth]{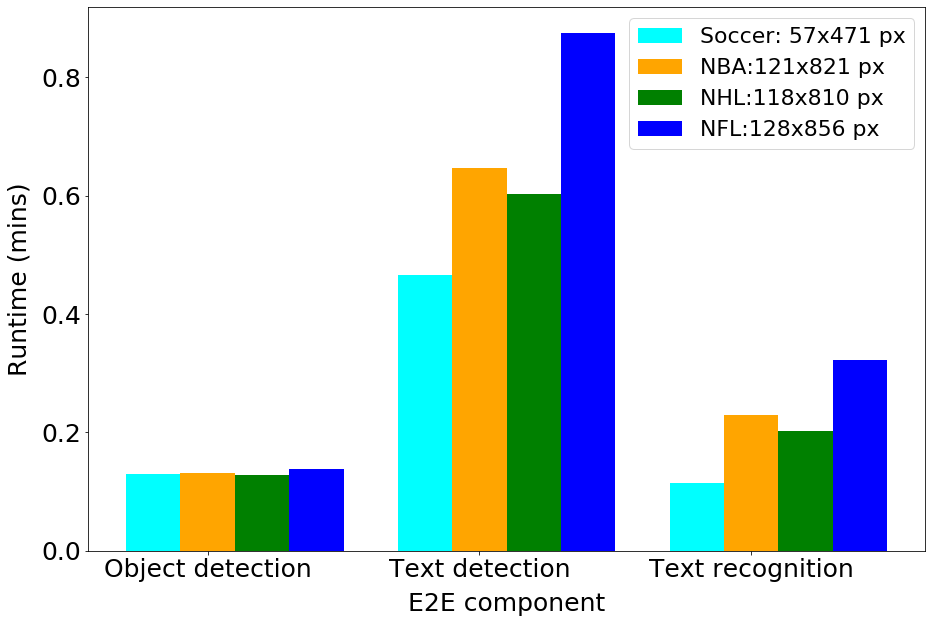}
        \vspace*{-6mm}
        \caption{Average runtime for processing one min long video clip from various sports domains on v100 gpu; legend represents clock types and average size}
        \label{fig:Runtime bargraphs}
    \end{subfigure}
    \caption{Computational Architecture and processing run time taken by different components on v100 GPU }
    \label{fig:CompArch and Process time}
    \vspace*{-4mm}
\end{figure*}

\section{System Architecture and runtime performance}

In this section, we describe the system architecture of our sports video understanding framework. Figure \ref{fig:computational-architecture} outlines the high-level architecture overview of the system.
\vspace*{-2mm}
\subsection{System Overview}
\label {subsec:sysoverview}
Both licensed and crawled sports videos (including YouTube) are processed through this computational architecture. \textit{Deep linked high recall} {player, action} triples are extracted from these sports video multimedia content,  apart from \textit{low recall} triples obtained from text. The video metadata (video url, title, description, video creation time) per produced video is asynchronously added to video information extraction queue (as shown in Figure \ref{fig:computational-architecture}, component 1). The queue is drained by a cluster of gpu enabled computational units which act as consumers. Each video event usually consists of a video of maximum length of 15 minutes. Longer full game videos are usually broken into smaller segments and playlist manifests are converted to messages, which can be parallely processed by different computational units in the cluster. Figure \ref{fig:computational-architecture} shows the different components which are described below:
\newline\textbf{Component 1, 2}: Downloads the video file/segment (mp4 or otherwise) using the video url to the local disk. In case of crawled videos (like YouTube) we need to extract the downloadable video url from the video player in the loaded html page. Multi-threaded ffmpeg (on cpu) decodes the video file to image sequences at a reduced frame rate (1/3rd) than the original (usually 25 or 30 fps), at a smaller resolution( 512 x 512 ) than the original (usually 720p or 1080p). The lower resolution, as well as the video chunk size (in case of long full game videos) is chosen such that for the given length of the video, they fit in gpu's native memory in a single batch prediction call to the trained game clock object detection model (specifically we train a SSD512 model).
\newline\textbf{Component 3}: Predictions of bounding box coordinates for each frame in input image sequence. Post prediction we drop the reduced image frame sequence from memory to keep the memory footprint low.
\newline\textbf{Component 4}: Post detection of clock bounding boxes, we scale the bounding box coordinates to original image resolution in the videos so that we can crop out natural sized clock images (again using ffmpeg decode). Post decode, we hold on to the cropped clock image objects (in memory) for every frame (at the reduced frame rate of processing) and send them to the text detection model to predict the text bounding boxes within the clock (as in Figure \ref{fig:computational-architecture} component 4).
\newline\textbf{Component 5}: The text regions identified within a clock are then passed to the text recognition model to extract textual strings.
\newline\textbf{Components 6}: In the post processing stage (as in Figure \ref{fig:computational-architecture} component 6), the extracted textual information from clock regions (teams, clock time and the current quarter) for the video are checked for consistency (across frames before and after) and filtered based on the knowledge constraints as explained in section \ref{sec:knowledge-constraints}.
\newline\textbf{Components 7,8,9,10}: The textual information from the post processing step is then aligned with the game play-by-play feed (as shown in Figure \ref{fig:frame-text-alignment}). Start and end time of the events are determined per sport (minimum time for of completion of a event given sport). The extracted information are then added to a vertical search index and as well fused with a larger knowledge graph to power rich sports cards.

 \begin{table}[]
 \caption{Sport leagues, YouTube channels and actions }
 \vspace{-4mm}
 \resizebox{\columnwidth}{!}{%
 \begin{tabular}{@{}lll@{}}
 \toprule
 Sport leagues    & YouTube channels                     & Detected actions                       \\ \midrule
 NBA              & NBA                                  & 3-pointer, 2-pointer, layup, slam-dunk \\
 NHL              & NBC Sports, MSG, TSN, NHL (Official) & goal, turnover, shot, penalty          \\
 NFL              & NBC Sports, NFL CBS, ESPN MNF, NFL   & tackle, touchdown, penalty, field goal \\
 Premiere League  & NBC Sports                           & goal, corner                           \\
 Champions League & B/R Football                         & goal, yellow card, wide                \\
 La Liga          & Bein Sports USA                      & goal                                   \\
 Europa league    & B/R Football                         & goal                                   \\ \bottomrule
 \end{tabular}%
 }
 \vspace{-3mm}
 \label{tab:coverage}
 \end{table}

 \begin{table}[]
 \caption{E2E pipeline component's runtime (mins) on sports videos with average duration of 8 mins}
 \vspace{-4mm}
 \resizebox{\columnwidth}{!}{%
 \begin{tabular}{@{}cccccccc@{}}
 \toprule
 GPU   & Parse video  & Object detection & Text detection & Text recognition  & Total runtime \\ \midrule
 p100  & 2.76                            & 1.34             & 4.88           & 1.64                                       & 10.62 \\
 v100  & 1.16                            & 1.05             & 5.21           & 1.73                                       & 9.15  \\ \bottomrule
 \end{tabular}%
 }
 \vspace{-5mm}
 \label{tab:avg runtime}
 \end{table}

\begin{table*}[]
\caption{End-to-end semantic classes text recognition performance by our models trained for NBA domain and models generalized to other sports subdomains on some YouTube videos. Videos 1 - 6 are from soccer and rest videos are from NFL and NHL sports. Each video has different style of clock. Models are never trained on clock samples from NFL or NHL sports.}
\vspace{-4mm}
\resizebox{\textwidth}{!}{%
\begin{tabular}{llllllllllllll}
\toprule
Video                          & Video 1 & Video 2 & Video 3 & Video 4 & Video 5                   & Video 6                   & Video 7 & Video 8 & Video 9 & Video 10 & Video 11         &         Video 12 & Video 13 \\
\midrule
 YouTube Id& ivHTSIzpzHo & FKbZYlcBNzQ & BN4lomASb1k & YycQcJKLBeM & aAGDem2XWi8 & LUvlSDwO2js & nMJVlo2h104 & T2DisWdsibw & dtoWs31-qj8 & RxKjW0Srk14 & xb4\_OYxoH0Q & xHBUkVqFHks & RJR3bAW4IIg \\
\midrule
Frames                         & 9330                        & 9600                        & 21750                       & 20370                       & 10230 & 13650 & 20220                       & 16260                       & 7590                        & 12210                        & 18180 & 15750                        & 15330                        \\
\midrule
Clocks                         & 5021                        & 6088                        & 13832                       & 12917                       & 7146  & 2168  & 18136                       & 15554                       & 6254                        & 6575                         & 17506 & 5831                         & 10053                        \\
\midrule
acc e-to-e (NBA models)   & 0.985                       & 0.571                       & 0.97        & 0.881                       &              0.972             &             0.971              & 0.894                       & 0.897                       & 0.969                       & 0.992                        &                0.48           & 0.981                        & 0.482                        \\
\midrule
acc e-to-e (Final models) & 0.991   & 0.923   & 0.983                       & 0.992                       &           0.983                &       0.973                    & 0.996                       & 0.994                       & 0.993                       & 0.993                        &          0.61                 & 0.992                        & 0.997   \\                    
\bottomrule
\end{tabular}%
}
\label{tab:NFL-NHL}
\vspace*{-3mm}
\end{table*}

\subsection{Runtime performance}
Table \ref{tab:coverage} shows a snapshot of the different YouTube channels we experimented on with our end-to-end processing pipeline, along with our licensed videos. NBA has a largely the variety of small sized clocks (similar to soccer leagues Premiere League, Champions's league, La Liga, or NHL) than NFL. Figure \ref{fig:CompArch and Process time}b shows the comparison of processing times for different leagues. Since clock object detection is done on a standard reduced size of the video (mentioned in section \ref{subsec:sysoverview}), prediction times are comparable across different leagues, while for text detections it is a function of the size of the clock (NFL being the largest) as they are performed on natural sized clocks. Finally bigger clock imply more text regions, thus increasing prediction time( as seen in NFL). Table \ref{tab:avg runtime} shows the average run times of different components across different sports on Nvidia P100 and V100 gpus both with 16GB of memory. Parse video (which is largely ffmpeg decode, running on cpus) time differently as they have different cpus. Most of the gpu bound components perform similarly in both gpu types.

\section{Training Results}

\begin{figure}[htp]
    \centering
    \begin{subfigure}[b]{\columnwidth}
        \centering
        \includegraphics[width=\textwidth]{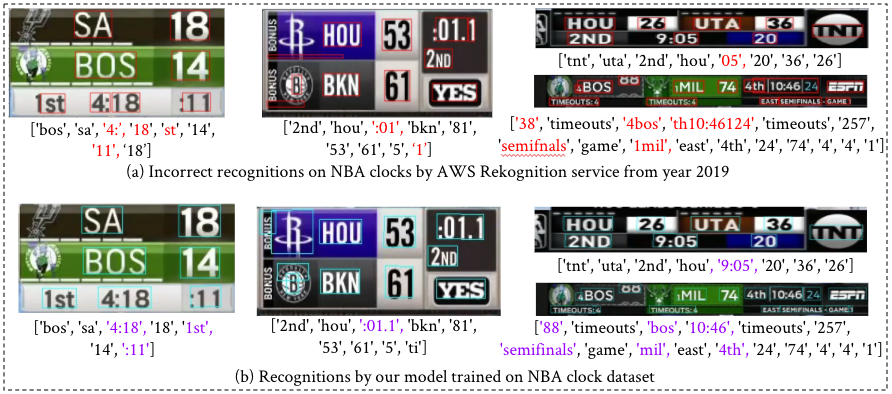}
        \captionsetup{size=scriptsize}
        \vspace*{-6mm}
        \caption{AWS Rekognition service vs our model trained on NBA clocks}
        \label{fig:AWSModel_errors}
     \end{subfigure}
     \begin{subfigure}[b]{\columnwidth} 
        \centering
        \includegraphics[width=\textwidth]{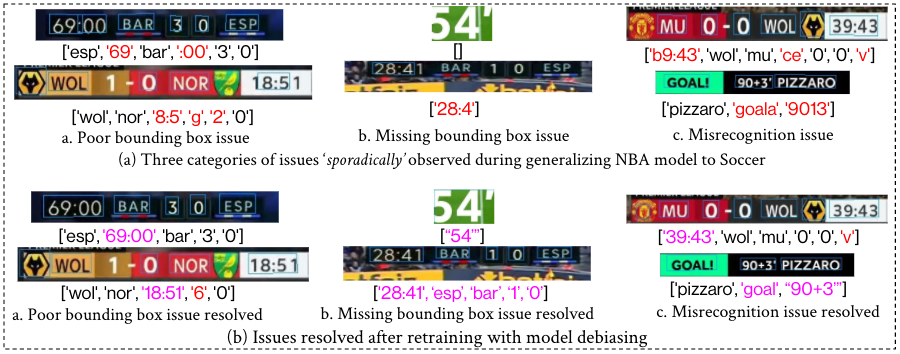}
        \captionsetup{size=scriptsize}
        \vspace*{-6mm}
        \caption{Text detection generalization issues across sub-domain and corrections post training}
        \label{fig:NBAModel_errors}
     \end{subfigure}
    \caption{Generalization errors and corrected predictions by transfer learned models}
    \label{fig:Clock issues}
\vspace{-4mm}
\end{figure}

\begin{table}[]
\caption{Performance of our models trained on the NBA dataset against the baseline}
\vspace{-4mm}
\begin{subtable}{0.4\textwidth}
\captionsetup{size=scriptsize}
\caption{Comparison of precision and recall over semantic classes and all classes for EAST model transfer learned on NBA clocks from pre-trained model as the baseline}
\vspace{-2mm}
\resizebox{\columnwidth}{!}{%
\begin{tabular}{lllllllll}
\toprule
IoU & P\_scb & P\_sc & R\_scb & R\_sc &P\_acb & P\_ac & R\_acb & R\_ac \\
\midrule
0.5 & 0.992 & 0.996 & 0.988 & 0.998 & 0.922 & 0.942 & 0.844 & 0.971 \\
0.6 & 0.983 & 0.995 & 0.988 & 0.998 & 0.889 & 0.934 & 0.839 & 0.971 \\
\textbf{0.7} & \textbf{0.88}  & \textbf{0.991} & \textbf{0.987} & \textbf{0.998} & \textbf{0.736} & \textbf{0.915} & \textbf{0.812} & \textbf{0.971} \\
0.8 & 0.504 & 0.958 & 0.977 & 0.998 & 0.381 & 0.851 & 0.692 & 0.968 \\
0.9 & 0.097 & 0.57  & 0.893 & 0.997 & 0.067 & 0.468 & 0.283 & 0.944 \\
\bottomrule
\end{tabular}
}
\label{tab:EASTvsBaseline}
\end{subtable}

\begin{subtable}{0.4\textwidth}
\captionsetup{size=scriptsize}
\caption{Comparison of accuracy over semantic classes and all classes for CRNN model trained on text from our NBA clock semantic classes and subset of synth90k jointly from clocks against baseline}
\vspace{-2mm}
\resizebox{\columnwidth}{!}{%
\begin{tabular}{lllllllll}
\toprule
IoU & Acc\_scb & Acc\_sc & Acc\_acb & Acc\_ac & C\_sc & I\_sc & C\_ac & I\_ac \\
\midrule
0.5 & 0.849 & 0.932 & 0.581 & 0.798 & 71821 & 5260 & 152761 & 38551 \\
0.6 & 0.855 & 0.933 & 0.593 & 0.804 & 71812 & 5168 & 152513 & 37199 \\
\textbf{0.7} & \textbf{0.877} & \textbf{0.936} & \textbf{0.622} & \textbf{0.814} & \textbf{71679} & \textbf{4898} & \textbf{151412} & \textbf{34519} \\
0.8 & 0.932 & 0.949 & 0.698 & 0.832 & 69599 & 3752 & 143864 & 28976 \\
0.9 & 0.962 & 0.97  & 0.743 & 0.854 & 41864 & 1315 & 81132  & 13863 \\
\bottomrule
\end{tabular}
}
\label{tab:CRNNvsBaseline}
\end{subtable}

\begin{subtable}{0.4\textwidth}
\captionsetup{size=scriptsize}
\caption{Comparison of precision and recall over semantic classes and all classes for e-to-e text detection and recognition system against baseline}
\vspace{-2mm}
\resizebox{\columnwidth}{!}{%
\begin{tabular}{lllllllll}
\toprule
IoU & P\_scb & P\_sc & R\_scb & R\_sc &P\_acb & P\_ac & R\_acb & R\_ac \\
\midrule
0.5 & 0.829 & 0.925 & 0.981 & 0.998 & 0.536 & 0.752 & 0.759 & 0.964 \\
0.6 & 0.826 & 0.924 & 0.981 & 0.998 & 0.527 & 0.751 & 0.756 & 0.964 \\
\textbf{0.7} & \textbf{0.762} & \textbf{0.923} & \textbf{0.979} & \textbf{0.998} & \textbf{0.457} & \textbf{0.745} & \textbf{0.729} & \textbf{0.964} \\
0.8 & 0.477 & 0.896 & 0.967 & 0.998 & 0.266 & 0.708 & 0.611 & 0.962 \\
0.9 & 0.093 & 0.539 & 0.85  & 0.997 & 0.05  & 0.399 & 0.226 & 0.935 \\
\bottomrule
\end{tabular}
}
\label{tab:E2EvsBaseline}
\end{subtable}
\label{tab:OurModelsvsBaseline}
\end{table}

\begin{table}[]
\caption{Performance of our models trained on the NBA\_Soc\_Augmented dataset}
\vspace{-4mm}
\resizebox{\columnwidth}{!}{%
\begin{tabular}{|l|rrrr|rr|rrrr|}
\toprule
 &
  \multicolumn{4}{c|}{\textbf{EAST}} &
  \multicolumn{2}{c|}{\textbf{CRNN}} &
  \multicolumn{4}{c|}{\textbf{End-to-End}} \\ \hline
\textbf{IoU} &
  \multicolumn{1}{l}{\textbf{P\_sc}} &
  \multicolumn{1}{l}{\textbf{R\_sc}} &
  \multicolumn{1}{l}{\textbf{P\_ac}} &
  \multicolumn{1}{l|}{\textbf{R\_ac}} &
  \multicolumn{1}{l}{\textbf{Acc\_sc}} &
  \multicolumn{1}{l|}{\textbf{Acc\_ac}} &
  \multicolumn{1}{l}{\textbf{P\_e2e\_sc}} &
  \multicolumn{1}{l}{\textbf{R\_e2e\_sc}} &
  \multicolumn{1}{l}{\textbf{P\_e2e\_ac}} &
  \multicolumn{1}{l|}{\textbf{R\_e2e\_ac}} \\ \hline
0.5 &
  0.995 &
  0.989 &
  0.949 &
  0.935 &
  0.937 &
  0.839 &
  0.93 &
  0.989 &
  0.796 &
  0.923 \\
\textbf{0.6} &
  \textbf{0.99} &
  \textbf{0.989} &
  \textbf{0.936} &
  \textbf{0.934} &
  \textbf{0.939} &
  \textbf{0.846} &
  \textbf{0.928} &
  \textbf{0.989} &
  \textbf{0.792} &
  \textbf{0.923} \\
0.7 &
  0.969 &
  0.989 &
  0.9 &
  0.931 &
  0.945 &
  0.858 &
  0.913 &
  0.989 &
  0.772 &
  0.921 \\
0.8 &
  0.869 &
  0.988 &
  0.767 &
  0.92 &
  0.955 &
  0.877 &
  0.825 &
  0.988 &
  0.673 &
  0.91 \\
0.9 &
  0.378 &
  0.973 &
  0.309 &
  0.823 &
  0.951 &
  0.882 &
  0.359 &
  0.972 &
  0.272 &
  0.804 \\
\bottomrule
\end{tabular}
}
\label{tab:evaluation_metrics}
\end{table}

\textbf{Text detector training}: We fine tune EAST \cite{zhou2017east}, pre-trained on  Incidental Scene Text dataset \cite{7333942} of ICDAR 13 and 15 (using Adam optimizer), for both semantic (time, quarter, teams) as well as other text regions (to avoid them being treated as hard negative samples) with a learning rate of 0.0001 and batch size of 12. Max and min text box sizes in training were 800px to 10px with minimum height/width ratio of 0.1 for 65 epochs on NBA data followed by 35 epochs on NBA\_Soc\_Augmented dataset. 
\newline\textbf{Text recognizer training}: We train a fresh CRNN \cite{shi2015endtoend} model with clock text regions pertaining to the semantic classes of time, quarter, teams (using SGD optimizer and a learning rate of 0.1/ dropout rate of  0.25) at a batch size of 256. Maximum label length is 23 with number of output timestamps set to 50. We trained CRNN \cite{shi2015endtoend} model with mix of NBA and Synth90k \cite{synth90k1,synth90k2} for 50 epochs, followed by fine tuning on NBA\_Soc\_Augmented (rich in special characters) for 32 epochs. 
\newline In Table \ref{tab:EASTvsBaseline} we report the performance of in-domain trained EAST (transfer learned from baseline) w.r.t baseline pre-trained EAST on Incidental Scene Text dataset \cite{7333942} of ICDAR 13 and 15. Combining it with tables \ref{tab:CRNNvsBaseline}, \ref{tab:E2EvsBaseline} and \ref{tab:evaluation_metrics} we compare end to end performance of training as we increase sports domains in our training datasets (namely NBA or NBA\_Soc\_Augmented). We do not train specifically on clocks obtained from american football (NFL) or ice-hockey (NHL) but find our model trained on the combined dataset (as described in section \ref{sec:dataaug}) to generalize well. \begin{math}P\_*,R\_*\end{math} denote precision and recall. Subscripts of \begin{math}sc,scb, ac,acb\end{math} signify only semantic classes, baseline model performance on semantic classes, all classes (i.e. text box of semantic text areas + other text) and baseline performance on all classes respectively. True positivies (TP), false positives (FP), and false negatives (FN) are calculated based on intersection over union (IoU) between predicted and ground truth boxes, at different thresholds of [0.5, 0.6, 0.7, 0.8, 0.9]. We find that even at a threshold of 0.8 the in-domain EAST model's precision and recall are very high. In Table \ref{tab:CRNNvsBaseline}, we compare the performance of an in-domain trained CRNN coupling with a baseline EAST model vs in-domain trained EAST model. It would be unfair to compare an in-the-wild trained CRNN model with an in-domain trained CRNN model as it does not have the full list of required characters to predict the semantic text successfully. Thus in Table \ref{tab:CRNNvsBaseline}, we test the robustness of an in-domain trained CRNN model's capability, under varying degrees of noise in text box predictions (based on IoU thresholds) to predict the entire semantic text string (i.e. full team names, time, quarter) correctly. We treat any partially correct recognition as a false positive. \begin{math}Acc\_*\end{math} signifies accuracy, while the subscripts have the same convention as mentioned above. Here also we see that under low IoU thresholds (0.5 in first row), the accuracy of in-domain trained CRNN model under supposedly noisy text box predictions, also is considerably high. In Table \ref{tab:E2EvsBaseline}, we measure the end to end performance of both EAST (pretrained baseline vs in-domain trained) and CRNN (in-domain trained), however accounting both systems in the precision and recall calculation. Across tables \ref{tab:EASTvsBaseline}, \ref{tab:CRNNvsBaseline}, \ref{tab:E2EvsBaseline} we find that at an operating range of EAST's predicted text box's IoU=0.7, we find good performance of the end to end systems on semantic classes. Table \ref{tab:evaluation_metrics} shows the component wise performance of both EAST and CRNN as well end to end on the augmented NBA\_Soc\_Augmented dataset as discussed in section \ref{sec:dataaug} and analyzed in section \ref{section:dataset}. We find that an operating range of IoU=0.6 we get high precision and recall of the semantic texts. Table \ref{tab:NFL-NHL} shows end to end performance on fresh videos scrapped from YouTube across different sports and comparing accuracy for models trained only over NBA data, versus NBA\_Soc\_Augmented data. The performance of video 11 is unusually low, primarily because we encounter a totally new styled clock. 

\section{Conclusion and future work} Figure \ref{fig:AWSModel_errors} and \ref{fig:NBAModel_errors} show how in-the-wild or not completely trained in domain detection and recognition models can impact precision and recall in a highly demanding mutimodal alignment scenario. We present an end to end system which scales both in terms of training and runtime. Prior works use play-by-play alignment over full game videos as a mechanism of generating training data, but our work showcases the intricacies of using a similar technique on highlight videos in a production environment. Finally we gather a rich dataset of fine grained shooting actions by running this pipeline in production, which we hope to contribute to the research community in future.

\clearpage



\bibliographystyle{ACM-Reference-Format}
\balance
\bibliography{ocr-bibtex}


\begin{thebibliography}{28}


\ifx \showCODEN    \undefined \def \showCODEN     #1{\unskip}     \fi
\ifx \showDOI      \undefined \def \showDOI       #1{#1}\fi
\ifx \showISBNx    \undefined \def \showISBNx     #1{\unskip}     \fi
\ifx \showISBNxiii \undefined \def \showISBNxiii  #1{\unskip}     \fi
\ifx \showISSN     \undefined \def \showISSN      #1{\unskip}     \fi
\ifx \showLCCN     \undefined \def \showLCCN      #1{\unskip}     \fi
\ifx \shownote     \undefined \def \shownote      #1{#1}          \fi
\ifx \showarticletitle \undefined \def \showarticletitle #1{#1}   \fi
\ifx \showURL      \undefined \def \showURL       {\relax}        \fi
\providecommand\bibfield[2]{#2}
\providecommand\bibinfo[2]{#2}
\providecommand\natexlab[1]{#1}
\providecommand\showeprint[2][]{arXiv:#2}

\bibitem[\protect\citeauthoryear{{Gerke}, {Müller}, and {Schäfer}}{{Gerke}
  et~al\mbox{.}}{2015}]%
        {Gerke}
\bibfield{author}{\bibinfo{person}{S. {Gerke}}, \bibinfo{person}{K. {Müller}},
  {and} \bibinfo{person}{R. {Schäfer}}.} \bibinfo{year}{2015}\natexlab{}.
\newblock \showarticletitle{Soccer Jersey Number Recognition Using
  Convolutional Neural Networks}.
\newblock  (\bibinfo{year}{2015}), \bibinfo{pages}{734--741}.
\newblock
\urldef\tempurl%
\url{https://doi.org/10.1109/ICCVW.2015.100}
\showDOI{\tempurl}


\bibitem[\protect\citeauthoryear{{Giancola}, {Amine}, {Dghaily}, and
  {Ghanem}}{{Giancola} et~al\mbox{.}}{2018}]%
        {SoccerNet}
\bibfield{author}{\bibinfo{person}{S. {Giancola}}, \bibinfo{person}{M.
  {Amine}}, \bibinfo{person}{T. {Dghaily}}, {and} \bibinfo{person}{B.
  {Ghanem}}.} \bibinfo{year}{2018}\natexlab{}.
\newblock \showarticletitle{SoccerNet: A Scalable Dataset for Action Spotting
  in Soccer Videos}.
\newblock  (\bibinfo{year}{2018}), \bibinfo{pages}{1792--179210}.
\newblock
\urldef\tempurl%
\url{https://doi.org/10.1109/CVPRW.2018.00223}
\showDOI{\tempurl}


\bibitem[\protect\citeauthoryear{Hu, Ma, Liu, Hovy, and Xing}{Hu
  et~al\mbox{.}}{2016}]%
        {hu2016harnessing}
\bibfield{author}{\bibinfo{person}{Zhiting Hu}, \bibinfo{person}{Xuezhe Ma},
  \bibinfo{person}{Zhengzhong Liu}, \bibinfo{person}{Eduard Hovy}, {and}
  \bibinfo{person}{Eric Xing}.} \bibinfo{year}{2016}\natexlab{}.
\newblock \showarticletitle{Harnessing deep neural networks with logic rules}.
\newblock \bibinfo{journal}{\emph{arXiv preprint arXiv:1603.06318}}
  (\bibinfo{year}{2016}).
\newblock


\bibitem[\protect\citeauthoryear{Hu, Yang, Salakhutdinov, Qin, Liang, Dong, and
  Xing}{Hu et~al\mbox{.}}{2018}]%
        {hu2018deep}
\bibfield{author}{\bibinfo{person}{Zhiting Hu}, \bibinfo{person}{Zichao Yang},
  \bibinfo{person}{Russ~R Salakhutdinov}, \bibinfo{person}{LIANHUI Qin},
  \bibinfo{person}{Xiaodan Liang}, \bibinfo{person}{Haoye Dong}, {and}
  \bibinfo{person}{Eric~P Xing}.} \bibinfo{year}{2018}\natexlab{}.
\newblock \showarticletitle{Deep generative models with learnable knowledge
  constraints}. In \bibinfo{booktitle}{\emph{Advances in Neural Information
  Processing Systems}}. \bibinfo{pages}{10501--10512}.
\newblock


\bibitem[\protect\citeauthoryear{Jaderberg, Simonyan, Vedaldi, and
  Zisserman}{Jaderberg et~al\mbox{.}}{2014}]%
        {synth90k1}
\bibfield{author}{\bibinfo{person}{Max Jaderberg}, \bibinfo{person}{Karen
  Simonyan}, \bibinfo{person}{Andrea Vedaldi}, {and} \bibinfo{person}{Andrew
  Zisserman}.} \bibinfo{year}{2014}\natexlab{}.
\newblock \showarticletitle{Synthetic Data and Artificial Neural Networks for
  Natural Scene Text Recognition}.
\newblock  (\bibinfo{year}{2014}).
\newblock


\bibitem[\protect\citeauthoryear{Jaderberg, Simonyan, Vedaldi, and
  Zisserman}{Jaderberg et~al\mbox{.}}{2016}]%
        {synth90k2}
\bibfield{author}{\bibinfo{person}{Max Jaderberg}, \bibinfo{person}{Karen
  Simonyan}, \bibinfo{person}{Andrea Vedaldi}, {and} \bibinfo{person}{Andrew
  Zisserman}.} \bibinfo{year}{2016}\natexlab{}.
\newblock \showarticletitle{Reading Text in the Wild with Convolutional Neural
  Networks}.
\newblock \bibinfo{journal}{\emph{International Journal of Computer Vision}}
  \bibinfo{volume}{116}, \bibinfo{number}{1} (\bibinfo{date}{jan}
  \bibinfo{year}{2016}), \bibinfo{pages}{1--20}.
\newblock


\bibitem[\protect\citeauthoryear{{Karatzas}, {Gomez-Bigorda}, {Nicolaou},
  {Ghosh}, {Bagdanov}, {Iwamura}, {Matas}, {Neumann}, {Chandrasekhar}, {Lu},
  {Shafait}, {Uchida}, and {Valveny}}{{Karatzas} et~al\mbox{.}}{2015}]%
        {7333942}
\bibfield{author}{\bibinfo{person}{D. {Karatzas}}, \bibinfo{person}{L.
  {Gomez-Bigorda}}, \bibinfo{person}{A. {Nicolaou}}, \bibinfo{person}{S.
  {Ghosh}}, \bibinfo{person}{A. {Bagdanov}}, \bibinfo{person}{M. {Iwamura}},
  \bibinfo{person}{J. {Matas}}, \bibinfo{person}{L. {Neumann}},
  \bibinfo{person}{V.~R. {Chandrasekhar}}, \bibinfo{person}{S. {Lu}},
  \bibinfo{person}{F. {Shafait}}, \bibinfo{person}{S. {Uchida}}, {and}
  \bibinfo{person}{E. {Valveny}}.} \bibinfo{year}{2015}\natexlab{}.
\newblock \showarticletitle{ICDAR 2015 competition on Robust Reading}. In
  \bibinfo{booktitle}{\emph{2015 13th International Conference on Document
  Analysis and Recognition (ICDAR)}}. \bibinfo{pages}{1156--1160}.
\newblock
\urldef\tempurl%
\url{https://doi.org/10.1109/ICDAR.2015.7333942}
\showDOI{\tempurl}


\bibitem[\protect\citeauthoryear{{Karatzas}, {Gómez}, {Nicolaou}, and
  {Rusiñol}}{{Karatzas} et~al\mbox{.}}{2018}]%
        {8395172}
\bibfield{author}{\bibinfo{person}{D. {Karatzas}}, \bibinfo{person}{L.
  {Gómez}}, \bibinfo{person}{A. {Nicolaou}}, {and} \bibinfo{person}{M.
  {Rusiñol}}.} \bibinfo{year}{2018}\natexlab{}.
\newblock \showarticletitle{The Robust Reading Competition Annotation and
  Evaluation Platform}. In \bibinfo{booktitle}{\emph{2018 13th IAPR
  International Workshop on Document Analysis Systems (DAS)}}.
  \bibinfo{pages}{61--66}.
\newblock
\urldef\tempurl%
\url{https://doi.org/10.1109/DAS.2018.22}
\showDOI{\tempurl}


\bibitem[\protect\citeauthoryear{Karatzas, Mestre, mas romeu, Nourbakhsh, and
  Roy}{Karatzas et~al\mbox{.}}{2011}]%
        {inproceedings}
\bibfield{author}{\bibinfo{person}{Dimosthenis Karatzas}, \bibinfo{person}{S.
  Mestre}, \bibinfo{person}{Juan mas romeu}, \bibinfo{person}{Farshad
  Nourbakhsh}, {and} \bibinfo{person}{P. Roy}.}
  \bibinfo{year}{2011}\natexlab{}.
\newblock \showarticletitle{ICDAR 2011 Robust Reading Competition - Challenge
  1: Reading Text in Born-Digital Images (Web and Email)}.
\newblock \bibinfo{journal}{\emph{Proceedings of the International Conference
  on Document Analysis and Recognition, ICDAR}}, \bibinfo{pages}{1485 -- 1490}.
\newblock
\urldef\tempurl%
\url{https://doi.org/10.1109/ICDAR.2011.295}
\showDOI{\tempurl}


\bibitem[\protect\citeauthoryear{Karatzas, Shafait, Uchida, Iwamura, Bigorda,
  Mestre, Mas, Mota, Almaz\`{a}n, and de~las Heras}{Karatzas
  et~al\mbox{.}}{2013}]%
        {10.1109/ICDAR.2013.221}
\bibfield{author}{\bibinfo{person}{Dimosthenis Karatzas},
  \bibinfo{person}{Faisal Shafait}, \bibinfo{person}{Seiichi Uchida},
  \bibinfo{person}{Masakazu Iwamura}, \bibinfo{person}{Lluis Gomez~i. Bigorda},
  \bibinfo{person}{Sergi~Robles Mestre}, \bibinfo{person}{Joan Mas},
  \bibinfo{person}{David~Fernandez Mota}, \bibinfo{person}{Jon~Almaz\`{a}n
  Almaz\`{a}n}, {and} \bibinfo{person}{Llu\'{\i}s~Pere de~las Heras}.}
  \bibinfo{year}{2013}\natexlab{}.
\newblock \showarticletitle{ICDAR 2013 Robust Reading Competition}. In
  \bibinfo{booktitle}{\emph{Proceedings of the 2013 12th International
  Conference on Document Analysis and Recognition}}
  \emph{(\bibinfo{series}{ICDAR '13})}. \bibinfo{publisher}{IEEE Computer
  Society}, \bibinfo{address}{USA}, \bibinfo{pages}{1484–1493}.
\newblock
\showISBNx{9780769549996}
\urldef\tempurl%
\url{https://doi.org/10.1109/ICDAR.2013.221}
\showDOI{\tempurl}


\bibitem[\protect\citeauthoryear{Khan, Lin, Tumrani, Wang, and Shao}{Khan
  et~al\mbox{.}}{2020}]%
        {ref2}
\bibfield{author}{\bibinfo{person}{Abdullah~Aman Khan},
  \bibinfo{person}{Haoyang Lin}, \bibinfo{person}{Saifullah Tumrani},
  \bibinfo{person}{Zheng Wang}, {and} \bibinfo{person}{Jie Shao}.}
  \bibinfo{year}{2020}\natexlab{}.
\newblock \showarticletitle{{Detection and localization of scorebox in long
  duration broadcast sports videos}}. In
  \bibinfo{booktitle}{\emph{International Symposium on Artificial Intelligence
  and Robotics 2020}}, \bibfield{editor}{\bibinfo{person}{Huimin Lu},
  \bibinfo{person}{Jože Guna}, {and} \bibinfo{person}{Yujie Li}} (Eds.),
  Vol.~\bibinfo{volume}{11574}. International Society for Optics and Photonics,
  \bibinfo{publisher}{SPIE}, \bibinfo{pages}{161 -- 172}.
\newblock
\urldef\tempurl%
\url{https://doi.org/10.1117/12.2575834}
\showURL{%
\tempurl}


\bibitem[\protect\citeauthoryear{Liao, Shi, and Bai}{Liao
  et~al\mbox{.}}{2018}]%
        {textbox}
\bibfield{author}{\bibinfo{person}{Minghui Liao}, \bibinfo{person}{Baoguang
  Shi}, {and} \bibinfo{person}{Xiang Bai}.} \bibinfo{year}{2018}\natexlab{}.
\newblock \showarticletitle{TextBoxes++: {A} Single-Shot Oriented Scene Text
  Detector}.
\newblock \bibinfo{journal}{\emph{CoRR}}  \bibinfo{volume}{abs/1801.02765}
  (\bibinfo{year}{2018}).
\newblock
\showeprint[arxiv]{1801.02765}
\urldef\tempurl%
\url{http://arxiv.org/abs/1801.02765}
\showURL{%
\tempurl}


\bibitem[\protect\citeauthoryear{{Liu} and {Bhanu}}{{Liu} and {Bhanu}}{2019}]%
        {Liu}
\bibfield{author}{\bibinfo{person}{H. {Liu}} {and} \bibinfo{person}{B.
  {Bhanu}}.} \bibinfo{year}{2019}\natexlab{}.
\newblock \showarticletitle{Pose-Guided R-CNN for Jersey Number Recognition in
  Sports}.
\newblock  (\bibinfo{year}{2019}), \bibinfo{pages}{2457--2466}.
\newblock
\urldef\tempurl%
\url{https://doi.org/10.1109/CVPRW.2019.00301}
\showDOI{\tempurl}


\bibitem[\protect\citeauthoryear{Liu, Anguelov, Erhan, Szegedy, Reed, Fu, and
  Berg}{Liu et~al\mbox{.}}{2015}]%
        {LiuAESR15}
\bibfield{author}{\bibinfo{person}{Wei Liu}, \bibinfo{person}{Dragomir
  Anguelov}, \bibinfo{person}{Dumitru Erhan}, \bibinfo{person}{Christian
  Szegedy}, \bibinfo{person}{Scott~E. Reed}, \bibinfo{person}{Cheng{-}Yang Fu},
  {and} \bibinfo{person}{Alexander~C. Berg}.} \bibinfo{year}{2015}\natexlab{}.
\newblock \showarticletitle{{SSD:} Single Shot MultiBox Detector}.
\newblock \bibinfo{journal}{\emph{CoRR}}  \bibinfo{volume}{abs/1512.02325}
  (\bibinfo{year}{2015}).
\newblock
\showeprint[arxiv]{1512.02325}
\urldef\tempurl%
\url{http://arxiv.org/abs/1512.02325}
\showURL{%
\tempurl}


\bibitem[\protect\citeauthoryear{Lu, Ting, Little, and Murphy}{Lu
  et~al\mbox{.}}{2013}]%
        {playerTrackMurphy}
\bibfield{author}{\bibinfo{person}{Wei-Lwun Lu}, \bibinfo{person}{Jo-Anne
  Ting}, \bibinfo{person}{J.J. Little}, {and} \bibinfo{person}{Kevin Murphy}.}
  \bibinfo{year}{2013}\natexlab{}.
\newblock \showarticletitle{Learning to Track and Identify Players from
  Broadcast Sports Videos}.
\newblock \bibinfo{journal}{\emph{IEEE transactions on pattern analysis and
  machine intelligence}}  \bibinfo{volume}{35} (\bibinfo{date}{07}
  \bibinfo{year}{2013}), \bibinfo{pages}{1704--16}.
\newblock
\urldef\tempurl%
\url{https://doi.org/10.1109/TPAMI.2012.242}
\showDOI{\tempurl}


\bibitem[\protect\citeauthoryear{Lucas, Panaretos, Sosa, Tang, Wong, Young,
  Ashida, Nagai, Okamoto, Yamamoto, Miyao, Zhu, Ou, Wolf, Jolion, Todoran,
  Worring, and Lin}{Lucas et~al\mbox{.}}{2005}]%
        {10.1007/s10032-004-0134-3}
\bibfield{author}{\bibinfo{person}{Simon~M. Lucas}, \bibinfo{person}{Alex
  Panaretos}, \bibinfo{person}{Luis Sosa}, \bibinfo{person}{Anthony Tang},
  \bibinfo{person}{Shirley Wong}, \bibinfo{person}{Robert Young},
  \bibinfo{person}{Kazuki Ashida}, \bibinfo{person}{Hiroki Nagai},
  \bibinfo{person}{Masayuki Okamoto}, \bibinfo{person}{Hiroaki Yamamoto},
  \bibinfo{person}{Hidetoshi Miyao}, \bibinfo{person}{Junmin Zhu},
  \bibinfo{person}{Wuwen Ou}, \bibinfo{person}{Christian Wolf},
  \bibinfo{person}{Jean-Michel Jolion}, \bibinfo{person}{Leon Todoran},
  \bibinfo{person}{Marcel Worring}, {and} \bibinfo{person}{Xiaofan Lin}.}
  \bibinfo{year}{2005}\natexlab{}.
\newblock \showarticletitle{ICDAR 2003 Robust Reading Competitions: Entries,
  Results, and Future Directions}.
\newblock \bibinfo{journal}{\emph{Int. J. Doc. Anal. Recognit.}}
  \bibinfo{volume}{7}, \bibinfo{number}{2–3} (\bibinfo{date}{July}
  \bibinfo{year}{2005}), \bibinfo{pages}{105–122}.
\newblock
\showISSN{1433-2833}
\urldef\tempurl%
\url{https://doi.org/10.1007/s10032-004-0134-3}
\showDOI{\tempurl}


\bibitem[\protect\citeauthoryear{Messelodi and Modena}{Messelodi and
  Modena}{2012}]%
        {s11042-011-0878-y}
\bibfield{author}{\bibinfo{person}{Stefano Messelodi} {and}
  \bibinfo{person}{Carla Modena}.} \bibinfo{year}{2012}\natexlab{}.
\newblock \showarticletitle{Scene text recognition and tracking to identify
  athletes in sport videos}.
\newblock \bibinfo{journal}{\emph{Multimedia Tools and Applications}}
  \bibinfo{volume}{63} (\bibinfo{date}{01} \bibinfo{year}{2012}),
  \bibinfo{pages}{1--25}.
\newblock
\urldef\tempurl%
\url{https://doi.org/10.1007/s11042-011-0878-y}
\showDOI{\tempurl}


\bibitem[\protect\citeauthoryear{Nady and Hemayed}{Nady and Hemayed}{2021}]%
        {DBLP:conf/visapp/NadyH21}
\bibfield{author}{\bibinfo{person}{Ahmed Nady} {and}
  \bibinfo{person}{Elsayed~E. Hemayed}.} \bibinfo{year}{2021}\natexlab{}.
\newblock \showarticletitle{Player Identification in Different Sports}.
\newblock  (\bibinfo{year}{2021}), \bibinfo{pages}{653--660}.
\newblock
\urldef\tempurl%
\url{https://doi.org/10.5220/0010341706530660}
\showDOI{\tempurl}


\bibitem[\protect\citeauthoryear{Piergiovanni and Ryoo}{Piergiovanni and
  Ryoo}{2018}]%
        {piergiovanni2018finegrained}
\bibfield{author}{\bibinfo{person}{AJ Piergiovanni} {and}
  \bibinfo{person}{Michael~S. Ryoo}.} \bibinfo{year}{2018}\natexlab{}.
\newblock \bibinfo{title}{Fine-grained Activity Recognition in Baseball
  Videos}.
\newblock
\newblock
\showeprint[arxiv]{1804.03247}~[cs.CV]


\bibitem[\protect\citeauthoryear{Ramanathan, Huang, Abu-El-Haija, Gorban,
  Murphy, and Fei-Fei}{Ramanathan et~al\mbox{.}}{2016}]%
        {ramanathan2016detecting}
\bibfield{author}{\bibinfo{person}{Vignesh Ramanathan},
  \bibinfo{person}{Jonathan Huang}, \bibinfo{person}{Sami Abu-El-Haija},
  \bibinfo{person}{Alexander Gorban}, \bibinfo{person}{Kevin Murphy}, {and}
  \bibinfo{person}{Li Fei-Fei}.} \bibinfo{year}{2016}\natexlab{}.
\newblock \bibinfo{title}{Detecting events and key actors in multi-person
  videos}.
\newblock
\newblock
\showeprint[arxiv]{1511.02917}~[cs.CV]


\bibitem[\protect\citeauthoryear{Shah, Schwartz, and Hovy}{Shah
  et~al\mbox{.}}{2020}]%
        {Shah_2020}
\bibfield{author}{\bibinfo{person}{Deven~Santosh Shah},
  \bibinfo{person}{H.~Andrew Schwartz}, {and} \bibinfo{person}{Dirk Hovy}.}
  \bibinfo{year}{2020}\natexlab{}.
\newblock \showarticletitle{Predictive Biases in Natural Language Processing
  Models: A Conceptual Framework and Overview}.
\newblock \bibinfo{journal}{\emph{Proceedings of the 58th Annual Meeting of the
  Association for Computational Linguistics}} (\bibinfo{year}{2020}).
\newblock
\urldef\tempurl%
\url{https://doi.org/10.18653/v1/2020.acl-main.468}
\showDOI{\tempurl}


\bibitem[\protect\citeauthoryear{Shi, Bai, and Belongie}{Shi
  et~al\mbox{.}}{2017}]%
        {seglink}
\bibfield{author}{\bibinfo{person}{Baoguang Shi}, \bibinfo{person}{Xiang Bai},
  {and} \bibinfo{person}{Serge~J. Belongie}.} \bibinfo{year}{2017}\natexlab{}.
\newblock \showarticletitle{Detecting Oriented Text in Natural Images by
  Linking Segments}.
\newblock \bibinfo{journal}{\emph{CoRR}}  \bibinfo{volume}{abs/1703.06520}
  (\bibinfo{year}{2017}).
\newblock
\showeprint[arxiv]{1703.06520}
\urldef\tempurl%
\url{http://arxiv.org/abs/1703.06520}
\showURL{%
\tempurl}


\bibitem[\protect\citeauthoryear{Shi, Bai, and Yao}{Shi et~al\mbox{.}}{2015}]%
        {shi2015endtoend}
\bibfield{author}{\bibinfo{person}{Baoguang Shi}, \bibinfo{person}{Xiang Bai},
  {and} \bibinfo{person}{Cong Yao}.} \bibinfo{year}{2015}\natexlab{}.
\newblock \bibinfo{title}{An End-to-End Trainable Neural Network for
  Image-based Sequence Recognition and Its Application to Scene Text
  Recognition}.
\newblock
\newblock
\showeprint[arxiv]{1507.05717}~[cs.CV]


\bibitem[\protect\citeauthoryear{Tian, Huang, He, He, and Qiao}{Tian
  et~al\mbox{.}}{2016}]%
        {ctpn}
\bibfield{author}{\bibinfo{person}{Zhi Tian}, \bibinfo{person}{Weilin Huang},
  \bibinfo{person}{Tong He}, \bibinfo{person}{Pan He}, {and}
  \bibinfo{person}{Yu Qiao}.} \bibinfo{year}{2016}\natexlab{}.
\newblock \showarticletitle{Detecting Text in Natural Image with Connectionist
  Text Proposal Network}.
\newblock \bibinfo{journal}{\emph{CoRR}}  \bibinfo{volume}{abs/1609.03605}
  (\bibinfo{year}{2016}).
\newblock
\showeprint[arxiv]{1609.03605}
\urldef\tempurl%
\url{http://arxiv.org/abs/1609.03605}
\showURL{%
\tempurl}


\bibitem[\protect\citeauthoryear{Xie, II, Jaques, Bailey, and Guturu}{Xie
  et~al\mbox{.}}{2021}]%
        {ref1}
\bibfield{author}{\bibinfo{person}{Dong Xie}, \bibinfo{person}{Arthur
  C.~Depoian II}, \bibinfo{person}{Lorenzo~E. Jaques},
  \bibinfo{person}{Colleen~P. Bailey}, {and} \bibinfo{person}{Parthasarathy
  Guturu}.} \bibinfo{year}{2021}\natexlab{}.
\newblock \showarticletitle{{Novel technique for broadcast footage overlay text
  recognition}}. In \bibinfo{booktitle}{\emph{Real-Time Image Processing and
  Deep Learning 2021}}, \bibfield{editor}{\bibinfo{person}{Nasser Kehtarnavaz}
  {and} \bibinfo{person}{Matthias~F. Carlsohn}} (Eds.),
  Vol.~\bibinfo{volume}{11736}. International Society for Optics and Photonics,
  \bibinfo{publisher}{SPIE}, \bibinfo{pages}{169 -- 175}.
\newblock
\urldef\tempurl%
\url{https://doi.org/10.1117/12.2588177}
\showURL{%
\tempurl}


\bibitem[\protect\citeauthoryear{{Yang}, {Yin}, {Yu}, {Karatzas}, and
  {Cao}}{{Yang} et~al\mbox{.}}{2017}]%
        {8270166}
\bibfield{author}{\bibinfo{person}{C. {Yang}}, \bibinfo{person}{X. {Yin}},
  \bibinfo{person}{H. {Yu}}, \bibinfo{person}{D. {Karatzas}}, {and}
  \bibinfo{person}{Y. {Cao}}.} \bibinfo{year}{2017}\natexlab{}.
\newblock \showarticletitle{ICDAR2017 Robust Reading Challenge on Text
  Extraction from Biomedical Literature Figures (DeTEXT)}. In
  \bibinfo{booktitle}{\emph{2017 14th IAPR International Conference on Document
  Analysis and Recognition (ICDAR)}}, Vol.~\bibinfo{volume}{01}.
  \bibinfo{pages}{1444--1447}.
\newblock
\urldef\tempurl%
\url{https://doi.org/10.1109/ICDAR.2017.235}
\showDOI{\tempurl}


\bibitem[\protect\citeauthoryear{Yu, Cheng, Ni, Wang, Zhang, and Yang}{Yu
  et~al\mbox{.}}{2018}]%
        {Yu_2018_CVPR}
\bibfield{author}{\bibinfo{person}{Huanyu Yu}, \bibinfo{person}{Shuo Cheng},
  \bibinfo{person}{Bingbing Ni}, \bibinfo{person}{Minsi Wang},
  \bibinfo{person}{Jian Zhang}, {and} \bibinfo{person}{Xiaokang Yang}.}
  \bibinfo{year}{2018}\natexlab{}.
\newblock \showarticletitle{Fine-Grained Video Captioning for Sports
  Narrative}.
\newblock  (\bibinfo{date}{June} \bibinfo{year}{2018}).
\newblock


\bibitem[\protect\citeauthoryear{Zhou, Yao, Wen, Wang, Zhou, He, and
  Liang}{Zhou et~al\mbox{.}}{2017}]%
        {zhou2017east}
\bibfield{author}{\bibinfo{person}{Xinyu Zhou}, \bibinfo{person}{Cong Yao},
  \bibinfo{person}{He Wen}, \bibinfo{person}{Yuzhi Wang},
  \bibinfo{person}{Shuchang Zhou}, \bibinfo{person}{Weiran He}, {and}
  \bibinfo{person}{Jiajun Liang}.} \bibinfo{year}{2017}\natexlab{}.
\newblock \bibinfo{title}{EAST: An Efficient and Accurate Scene Text Detector}.
\newblock
\newblock
\showeprint[arxiv]{1704.03155}~[cs.CV]


\end{thebibliography}

    

\end{document}